\documentclass[twocolumn]{aastex62}

\usepackage{CJKutf8} 

\received{November 16, 2018}
\revised{January 18, 2019}
\accepted{ January 27, 2019 }
 
\submitjournal{ApJ}

\begin{document}
\begin{CJK}{UTF8}{bsmi}
 
\title{Forward Modeling of \textit{SDO}/AIA and X-Ray Emission from a Simulated Flux Rope Ejection}

\author[0000-0003-2875-7366]{Xiaozhou Zhao (趙小舟) }
\affiliation{Key Laboratory of Dark Matter and Space Astronomy, Purple Mountain Observatory, Chinese Academy of Sciences,\\ 210008 Nanjing, China; zhaoxz@pmo.ac.cn, wqgan@pmo.ac.cn}
\affiliation{Centre for mathematical Plasma Astrophysics, Department of Mathematics, KU Leuven, Celestijnenlaan 200B, 3001 Leuven, Belgium; rony.keppens@kuleuven.be, tom.vandoorsselaere@kuleuven.be}
\affiliation{University of Chinese Academy of Sciences, Beijing 100049, PR China.}

\author[0000-0002-7153-4304]{Chun Xia}
\affiliation{School of Physics and Astronomy, Yunnan University,\\ 650050 Kunming, China; chun.xia@ynu.edu.cn}

\author[0000-0001-9628-4113]{Tom Van Doorsselaere}
\affiliation{Centre for mathematical Plasma Astrophysics, Department of Mathematics, KU Leuven, Celestijnenlaan 200B, 3001 Leuven, Belgium; rony.keppens@kuleuven.be, tom.vandoorsselaere@kuleuven.be}

\author[0000-0003-3544-2733]{Rony Keppens}
\affiliation{Centre for mathematical Plasma Astrophysics, Department of Mathematics, KU Leuven, Celestijnenlaan 200B, 3001 Leuven, Belgium; rony.keppens@kuleuven.be, tom.vandoorsselaere@kuleuven.be}

\author{Weiqun Gan }
\affiliation{Key Laboratory of Dark Matter and Space Astronomy, Purple Mountain Observatory, Chinese Academy of Sciences,\\ 210008 Nanjing, China; zhaoxz@pmo.ac.cn, wqgan@pmo.ac.cn}

\begin{abstract}

We conduct forward-modeling analysis based on our 2.5 dimensional magnetohydrodynamics (MHD) simulation of magnetic flux rope (MFR) formation and eruption driven by photospheric converging motion. The current sheet (CS) evolution during the MFR formation and eruption process in our MHD simulation can be divided into four
stages. The current sheet (CS) evolution during the MFR formation and eruption process in our MHD simulation can be divided into four stages. The first stage shows the CS forming and gradually lengthening. Resistive instabilities that disrupt the CS mark the beginning of the second stage. Magnetic islands disappear in the third stage and reappear in the fourth stage. Synthetic images and light curves of the seven \textit{Solar Dynamics Observatory}/Atmospheric Imaging Assembly (AIA) channels, i.e., $94\,\mathrm{\AA}$, $131\,\mathrm{\AA}$, $171\,\mathrm{\AA}$, $193\,\mathrm{\AA}$, $211\,\mathrm{\AA}$, $304\,\mathrm{\AA}$, and $335\,\mathrm{\AA}$, and the $3-25\,\mathrm{keV}$ thermal X-ray are obtained with forward-modeling analysis. The loop-top source and the coronal sources of the soft X-ray are reproduced in forward modeling. The light curves of the seven \textit{SDO}/AIA channels start to rise once resistive instabilities develop. The light curve of the $3-25\,\mathrm{keV}$ thermal X-ray starts to go up when the reconnection rate reaches one of its peaks. Quasiperiodic pulsations (QPPs) appear twice in the \textit{SDO}/AIA $171\,\mathrm{\AA}$, $211\,\mathrm{\AA}$, and $304\,\mathrm{\AA}$ channels, corresponding to the period of chaotic (re)appearance and CS-guided displacements of the magnetic islands. QPPs appear once in the \textit{SDO}/AIA $94\,\mathrm{\AA}$ and $335\,\mathrm{\AA}$ channels after the disruption of the CS by resistive instabilities and in the $193\,\mathrm{\AA}$ channel when the chaotic motion of the magnetic islands reappears.

\end{abstract}
 

\section{Introduction  } \label{sec:1}

Coronal mass ejections (CMEs) are large-scale ejections of mass and magnetic flux from the solar corona into space. The CME progenitor is believed to be a helical magnetic flux rope (MFR) buried in the low corona. The standard CSHKP scenario of flares \citep{Carmichael1964,Sturrock1966,Hirayama1974,KoppPneuman1976} based on observations is a well-established phenomenological model that depicts the overall evolution of eruptive flares/CMEs. A current sheet (CS) develops between the flare arcade and the CME bubble \citep{ForbesActon1996,Hu2000ApJ,LinForbes2000,Takahashi2017}. CS fragmentation, accompanied by the production of multiple magnetic islands of different sizes, is reported in both observations \citep{Karlick2010,Takasao2012,Takasao2016} and simulations \citep{Ni2012,Guo2013,Guidoni2016,Mei2017} during the flare/CME process.  Recently, \citet{Ruan2018arXiv} argued for a novel ingredient in the standard CSHKP model, namely a Kelvin-Helmholtz interaction at the loop top in the underlying flare loops. Their simulations studied the evaporation process in flare loops, in isolation of the overarching CS.

The CS structure, the reconnection inflow and outflow, and the plasmoids have been observed several times. The reconnection CS predicted by MFR models of CMEs was observed, e.g., by \citet{Ciaravella2002}. \citet{Asai2004} examined the downflow motions above flare loops that are correlated with reconnection outflows, noting the deceleration of the downflows at the top of the postflare loops. \citet{Ko2003} saw blobs steadily flowing along the CS. \citet{Song2012} statistically investigated the morphology of the CS and blob dynamics. \citet{Takasao2012} reported the simultaneous observation of magnetic reconnection inflow and outflow in a flare by \textit{Solar Dynamics Observatory SDO}/Atmospheric Imaging Assembly (AIA), and the coalescence of the plasmoids.

Quasirepetitive patterns in the emission of solar flares, i.e., the periodic intensity increase and decrease of flare light curves, are called quasiperiodic pulsations (or QPPs) \citep{VanDoorsselaere2016SoPh,McLaughlin2018}. The physical mechanism responsible for the generation of QPPs is still under debate.  It is possible that QPPs are triggered by an external oscillatory inflow that modulates the reconnection periodically \citep{Nakariakov2006,Gruszecki2011,Nakariakov2011,Santamaria2015}. Another possibility for QPPs is the inherent periodicity of the reconnection during the flare \citep{Longcope2007,Takasao2015}.
QPPs may also be generated by the waves reflected several times in the flaring loop \citep{Dolla2012,Fang2015,Yuan2015}, because waves are considered to be as a candidate transport mechanism for the flare energy from the X-point location to the footpoints \citep{Fletcher2008,Russell2013}. Most recently, \citet{Ruan2018arXiv} demonstrated that the Kelvin-Helmholtz turbulence associated with interacting evaporation flows in the postflare loop can indeed cause periodic variations in light curves, in agreement with observations.

Physical quantities (e.g., density, temperature, and pressure) obtained in numerical simulations cannot be directly compared with observational data because only spectral-line profiles (integrated over space, time, and wavelength, depending on the instrument) are obtained in observations. To compare simulations with observations, it is necessary to convert simulation data to pseudo-emission and create artificial observations, a process called forward modeling \citep{VanDoorsselaere2016}. The AIA \citep{Lemen2012} on board the \textit{SDO} \citep{Pesnell2012} provides full disk images of the Sun with seven extreme ultraviolet (EUV) bandpasses centered on specific lines: Fe XVIII ($94\,\mathrm{\AA}$), Fe VIII, XXI ($131\,\mathrm{\AA}$), Fe IX ($171\,\mathrm{\AA}$), Fe XII, XXIV ($193\,\mathrm{\AA}$), Fe XIV ($211\,\mathrm{\AA}$), He II ($304\,\mathrm{\AA}$), and Fe XVI ($335\,\mathrm{\AA}$). To compare MHD simulations with AIA observations, MHD simulation results should be converted to AIA emissions in various bandpasses by forward modeling.

Efforts have been put into bridging the gap between models and observations. \citet{Gruszecki2012} simply supposed that the thermal emission is optically thin and used the essential fact that optically thin emission is always density-squared dependent. \citet{Pagano2014} produced synthesized \textit{SDO}/AIA observation from the simulated MFR ejection. Synthetic images of three-dimensional MHD simulations are obtained by \citet{Xia2014ApJ,Xia2017AA} and \citet{Xia2016ApJ}. There are several tools available for forward modeling, e.g., \textit{FORWARD SolarSoft IDL package}, which is developed for the synthesis of multi-wavelength coronal data from a given coronal model \citep{Gibson2015,Gibson2016}, and \textit{GX Simulator}, which can produce multiwavelength images and spectra of radio and X-ray emissions from a given coronal model \citep{Nita2015}. \textit{FoMo}, developed at the KU Leuven (Belgium), is another useful tool for forward modeling \citep{VanDoorsselaere2016}. The MHD simulation code \textit{MPI-AMRVAC} \citep{Keppens2012JCoPh,Porth2014ApJS,Xia2018ApJS} can also transfer the three-dimensional MHD studies to synthetic images.

In this paper, we conduct forward-modeling analysis based on our 2.5 dimensional simulation of MFR formation and eruption. Section \ref{sec:2} describes the numerical model and illustrates the numerical results. Section \ref{sec:3} explains the forward-modeling technique, which goes beyond the capabilities of \textit{FoMo}, as we account for effects like partial ionization, absorption, etc. Section \ref{sec:4} shows the synthetic images and discusses the light curves obtained in the forward modeling. The paper is summarized in Section \ref{sec:5}.

\section{Numerical Model   }\label{sec:2}

\citet{Zhao2017} simulated the formation and eruption of an MFR driven by photospheric converging motion in a chromosphere-transition-corona setup, including the effects of radiative cooling, anisotropic thermal conduction along the magnetic field lines, gravitational stratification, resistivity, and viscosity. Our simulation in this paper will use the same physical setup, but contrary to our previous study, we include the thermal flux saturation effect \citep{Cowie1977ApJ} into the numerical model of \citet{Zhao2017}. During the impulsive reconnection phase, steep temperature gradients may develop and the classical thermal conduction may saturate. On the one hand, the thermal flux saturation effect makes the solar atmosphere model more realistic in this work, and, on the other hand, the numerical computation is more stable, because the steepness of the temperature gradients is limited. The present simulation is able to follow the eruption stage for a much longer time. After including the thermal flux saturation effect, we recomputed, keeping all the numerical setups, i.e., the numerical methods, the initial and boundary conditions, and the normalization units of MHD equations, the same as that in \citet{Zhao2017}. The simulation presented here was realized by the \textit{MPI-AMRVAC} simulation toolkit.

\subsection{Global Evolution } \label{sec:2.1}
\begin{figure*}[ht!]
	\epsscale{1.2}
	\plotone{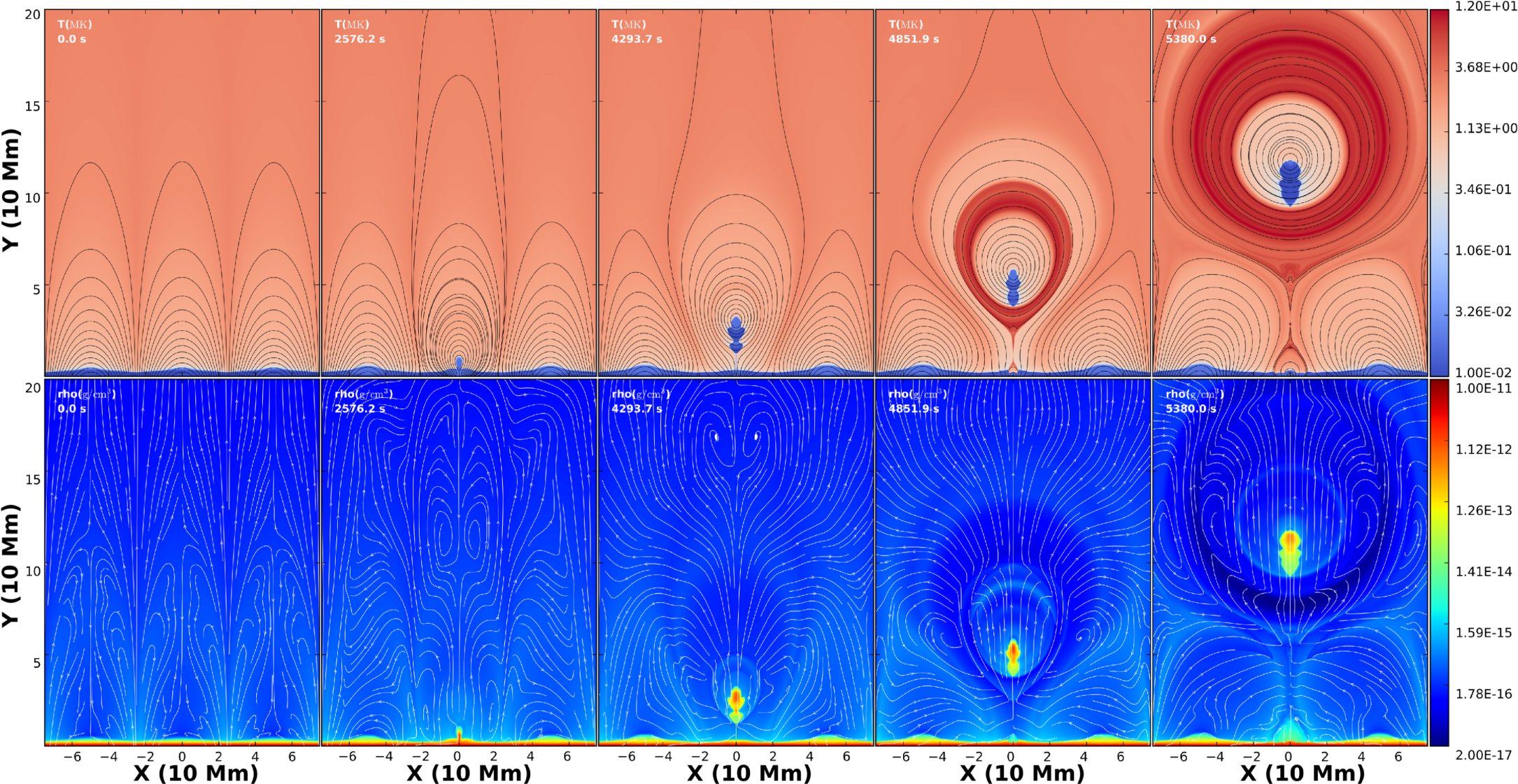}  
	
	\caption{Evolution of temperature (upper) and density (lower) in the {\it x-y} plane, indicating the MFR formation and eruption. Magnetic field lines (upper) and velocity stream lines (lower) are overplotted. The axis scales are in units of $10\,\mathrm{Mm}$.\label{fig1}}
\end{figure*}

The snapshots in Figure \ref{fig1} capture the evolution of the temperature distribution in the {\it x-y} plane with magnetic field lines overlaid (upper panels) and the evolution of the mass density distribution with streamlines (lower panels). Each snapshot, the size of which is $[-75\,\mathrm{Mm},75\,\mathrm{Mm}]\times[0\,\mathrm{Mm} ,200\,\mathrm{Mm}]$, does not correspond to the entire simulation box but covers a part of it, where the formation and eruption of the MFR occur. The photospheric converging motion brings magnetic field lines of opposite polarities to the magnetic reversal line and drives magnetic reconnection there, leading to the formation of an MFR at $t\sim2112.5\,\mathrm{s}$. The newly formed MFR rises into the corona, lifting mass from the chromosphere into the corona, giving rise to a prominence. We see from Figure \ref{fig1} that chromospheric mass has been lifted into the corona at $t\sim4293.7\,\mathrm{s}$. A CS develops below the MFR. The CS extends in length in the vertical direction as the MFR rises. Once the aspect ratio of the CS exceeds a critical value, resistive instabilities start, and multiple magnetic islands of different sizes are formed. For a more detailed discussion of this process, consult \citet{Zhao2017}. The hot reconnection outflow hits the bottom of the MFR, then mass and heat are transported along the magnetic field lines around the MFR, forming a layered structure of the temperature distribution of the MFR. At $t\sim4851.9\,\mathrm{s}$, the outer layer of the MFR is hot, with a temperature of about $10^{7}\,\mathrm{K}$, while the temperature of the region surrounding the prominence is less than a million Kelvin, about $8\times10^{5}\,\mathrm{K}$. At $t\sim5380.0\,\mathrm{s}$, the temperature of the region surrounding the prominence is almost the same as before, while the outermost layer of the MFR is about $6\times10^{6}\,\mathrm{K}$. An intermediate layer with a temperature of about $1.2\times10^{7}\,\mathrm{K}$ is hotter than the region surrounding the prominence and the outermost layer of the MFR. The thermal conduction between different layers is negligible due to the anisotropy of thermal conduction, thus the heat is confined in each layer. The layered temperature structure indicates that the temperature of the reconnection outflow varies with time because the inner layers are formed earlier than the outer layers. Figure \ref{fig2} is the time-distance diagram of mass density along the line $x=0$. The solid black line represents the height of the center of the MFR. The first vertical solid line indicates the instant when resistive instabilities start, which is $t=4304.0\,\mathrm{s}$. The acceleration of the MFR increases once resistive instabilities start, which is discussed in \citet{Zhao2017}.

\subsection{CS Evolution  }\label{sec:2.2}

\begin{figure*}[ht!]
	\epsscale{1.2}
	\plotone{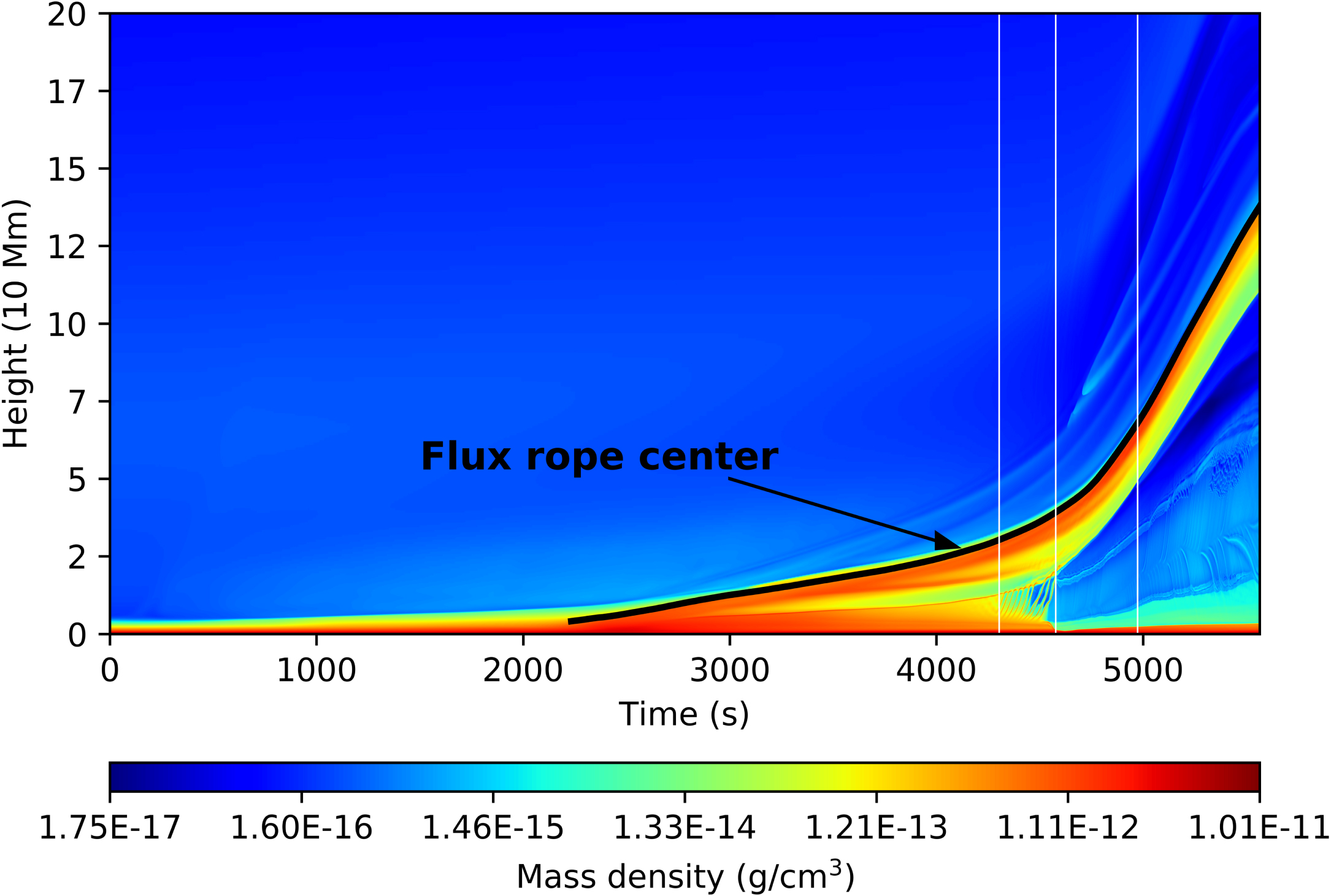}  	
	\caption{Time-distance diagram of mass density along the line $x=0$. The first vertical solid line indicates the instant $t=4304.0\,\mathrm{s}$ when resistive instabilities start in the CS. The small-scale island dynamics is visible below the flux rope. The second vertical solid line indicates the instant $t=4577.1\,\mathrm{s}$ when the hot CS stage of the CS evolution begins. The third vertical solid line indicates the instant $t=4973.0\,\mathrm{s}$ when the dynamic hot CS stage of the CS evolution starts. The solid black line represents the height of the flux rope center.\label{fig2}} 
\end{figure*}

\begin{figure*}[ht!]
	\epsscale{1.26}
	\plotone{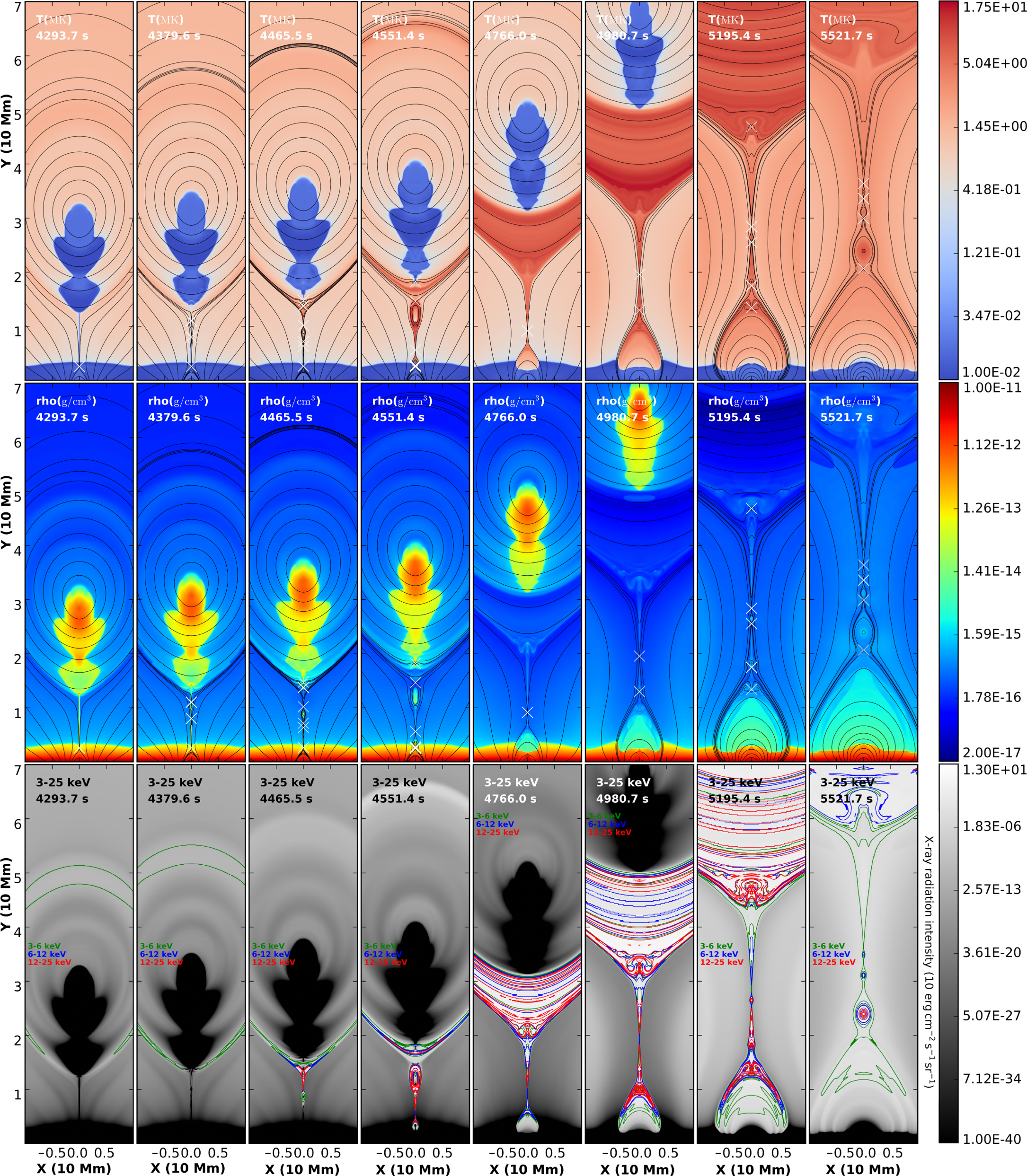}  
	
	\caption{Evolution of temperature (upper panels), density (middle panels), and $3-25\,\mathrm{keV}$ X-ray intensity (bottom panels) in the {\it x-y} plane, indicating the CS evolution. In the upper and middle panels, the solid black lines and the white crosses  represent the magnetic field lines and the X-points, respectively. The contours of the $3-6\,\mathrm{keV}$, $6-12\,\mathrm{keV}$, and $12-25\,\mathrm{keV}$ X-ray intensities are plotted in the bottom panels with contour levels of  $5\times 10^{n\Delta-6}I_{max}$, where $n=0,1,2,3$, $4$, $\Delta=(6-\lg 5)/4$, and $I_{max}$ represents the maximum value of the radiation intensity of each wavelength at each instant.\label{fig3}}
	
\end{figure*}

The simulation by \citet{Zhao2017} stops at $t\sim4500\,\mathrm{s}$ while the simulation in this work lasts up to $t\sim5500\,\mathrm{s}$, thus the CS evolution process is more complete. Figure \ref{fig3} shows the CS evolution with temperature (upper panels), mass density (middle panels), and X-ray intensity (lower panels) plotted. The X-ray intensity will be discussed in Section \ref{sec:3} and Section \ref{sec:4}. We divide the CS evolution into four stages in this work, and the starting and ending time, the temperature of the CS, the density of the CS and number of the magnetic islands at each stage of the CS evolution are listed in Table \ref{tbl1}. The first X-point in this simulation appears at $t\sim2112.5\,\mathrm{s}$, indicating that the CS is created then, and the CS growth stage begins. The CS grows in length as the MFR rises. The first column of Figure \ref{fig3} indicates the snapshots at a late time of the CS growth stage at $4293.7\,\mathrm{s}$. The density of the CS is higher than the ambient coronal plasma, while the temperature is lower than the ambient coronal plasma. There is only one X-point (white cross) in the CS, which is buried in the chromosphere. The CS growth stage of the evolution is to the left of the first vertical solid line in Figure \ref{fig2}. The CS connects the prominence and the chromosphere in the CS growth stage, and mass motion in the CS is not evident in Figure \ref{fig2}. The dynamic growth stage starts at $t\sim4304.0\,\mathrm{s}$ when resistive instabilities are ignited. The second, third and fourth columns of Figure \ref{fig3} indicate the second-stage evolution of the CS at $4379.6\,\mathrm{s}$, $4465.5\,\mathrm{s}$, and $4551.4\,\mathrm{s}$, respectively. Multiple X-points (white crosses) and magnetic islands appear at this stage. The magnetic islands move and transport the dense and cool mass in the CS upward or downward, making the density of the CS decrease. In the dynamic growth stage, the CS is heated from $T\sim10^{4}\,\mathrm{K}$ to $T\sim10^{7}\,\mathrm{K}$ by the reconnection while the density decreases due to the motion of the magnetic islands. The prominence detaches from the bottom of the MFR, and the CS is no longer connected to the prominence in this stage. In Figure \ref{fig2}, the dynamic growth stage of the evolution is represented by the region between the first and second vertical solid lines, where the bulk motion of the individual islands is clearly shown. The hot CS stage starts at $t\sim4577.1\,\mathrm{s}$. The fifth column of Figure \ref{fig3} indicates the hot CS stage of the evolution at $t=4766.0\,\mathrm{s}$. There are very few magnetic islands in the hot CS stage. In the hot CS stage, some parts of the CS turn to the exhaust outflow region, and the length of the body of the CS, i.e. the part where the electric current concentrates, decreases, leading to the disappearance of magnetic islands. In Figure \ref{fig2}, the hot CS stage is represented by the region lying between the second and third vertical solid lines, where the motion of the magnetic islands is not obvious. The temperature and density of the CS are close to the ambient coronal plasma in this stage, which are about $6\times10^{6}\,\mathrm{K}$ and $8.2\times 10^{-16} \,\mathrm{g\cdot cm^{-3}}$, respectively. The dynamic hot CS stage of the evolution starts at $\sim4973.0\,\mathrm{s}$. This stage is indicated in the sixth, seventh, and eighth columns of Figure \ref{fig3}, i.e., snapshots at $t=4980.7\,\mathrm{s}$, $t=5195.4\,\mathrm{s}$, and $t=5521.7\,\mathrm{s}$, respectively, and the right side of the third vertical solid line in Figure \ref{fig2}. As the MFR rises, the CS grows in length, and resistive instabilities start again in this stage. Multiple X-points and magnetic islands can be seen in the sixth, seventh, and eighth columns in Figure \ref{fig3}. The trajectories of the magnetic islands can be seen in the CS region in Figure \ref{fig2}, which is located on the right side of the third vertical solid line and below the prominence and the MFR. The density and temperature in this stage remain almost the same as those in the hot CS stage.

\begin{table*}[t]
	\begin{center}
		\caption{Current Sheet Evolution.\label{tbl1}}
		\begin{tabular}{r|r |r| r| r|r  }
			\tableline\tableline
			\centering
			Stage  & Name of Stage  & Time & Temperature & Density  &Number of Islands\\
			\tableline
			First  &CS growth stage & $2112.5\,\mathrm{s}-4304.0\,\mathrm{s}$ & $3\times10^{4}\,\mathrm{K}$ & $1.64\times10^{-13}\,\mathrm{g\cdot cm^{-3}}$ & Not appearing\\
			\tableline
			Second  &Dynamic growth stage  &$4304.0\,\mathrm{s}-4577.1\,\mathrm{s}$ & $3\times10^{4}-7\times10^{6}\,\mathrm{K}$ &  $2.0\times10^{-15}-3.0\times10^{-13}\,\mathrm{g\cdot cm^{-3}}$  & 1-13 \\
			\tableline
			Third  &Hot CS stage  &$4577.1\,\mathrm{s}-4973.0\,\mathrm{s}$ & $6\times10^{6}\,\mathrm{K}$ & $8.2\times 10^{-16} \,\mathrm{g\cdot cm^{-3}}  $  & 0\\
			\tableline
			Fourth  &Dynamic hot CS stage  &$4973.0\,\mathrm{s}-5564.7\,\mathrm{s}$ & $5\times10^{6}\,\mathrm{K}$ & $8.2\times 10^{-16} \,\mathrm{g\cdot cm^{-3}}$  & 1-25\\
			
			\tableline
		\end{tabular}
	\end{center}
\end{table*}

\subsection{Magnetic Reconnection Rate  }\label{sec:2.3}

The magnetic reconnection rate $E_{rec}$ is the rate of flux transport across the X-point \citep{Biskamp1997,PriestForbes2000,Priest2016ASSL}, which is calculated as follows: 
\begin{equation}
E_{rec}=\left |-\frac{1}{c} \frac{\mathrm{d} A}{\mathrm{d} t} \right |.\label{eq1}
\end{equation}
Here, the flux function $A$ determines the magnetic field in the {\it x-y} plane by $B_{x}=-\partial A/\partial y$ and $B_{y}=\partial A/\partial x$, which is the two-dimensional counterpart of the vector potential in three-dimensional space. The appearance of the light speed $c$ in the above equation makes the reconnection rate in the same dimension as the electric field in Gaussian units.
It should be noted that the reconnection rate so defined represents the value of the electric field $\mathbf{E}_{rec}$ measured in the observer's frame of reference, while the electric field $\mathbf{E}^{\prime}$ measured in the comoving frame of reference of the X-point is given by Ohm's law as $\mathbf{E}^{\prime}=\eta\mathbf{j}$. They are connected by a Galilean transformation \citep{Landau1975} as follows:
\begin{equation}
\mathbf{E}^{\prime}=\mathbf{E}_{rec}+\frac{1}{c}\mathbf{v}\times \mathbf{B},\label{eq2}
\end{equation}
where the magnetic field $\mathbf{B}$ is considered not to change in the Galilean transformation \citep{Biskamp1997} and $\mathbf{v}$ is the velocity of the X-point relative to the observer. So, the reconnection rate of a two-dimensional system, i.e., the electric field measured in the observer's frame of reference, is  
\begin{equation}
E_{rec}=\left |\eta j_{z}+  \frac{v_{y}B_{x}}{c}  -\frac{v_{x}B_{y}}{c}   \right |.\label{eq3}
\end{equation}

\begin{figure*}[ht!]
	\epsscale{1.16}
	\plotone{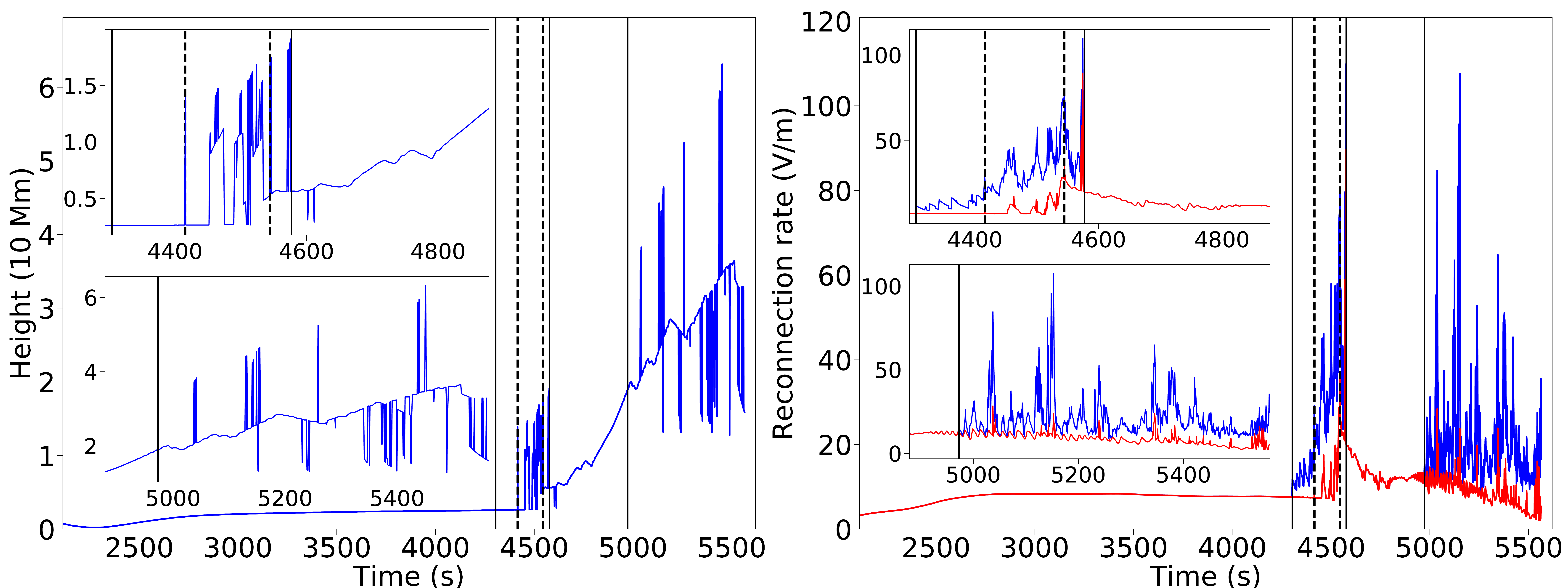}  
	
	\caption{The height of the principal X-point is plotted on the left. The global reconnection rate (blue line) and the reconnection rate at the principal X-point (red line) are plotted on the right. The first vertical solid line indicates the instant $t=4304.0\,\mathrm{s}$ when the resistive instabilities start in the CS. The second vertical solid line indicates the instant $t=4577.1\,\mathrm{s}$ when the hot CS stage of the CS evolution begins. The third vertical solid line indicates the instant $t=4973.0\,\mathrm{s}$ when the dynamic hot CS stage of the CS evolution starts. The first vertical dashed line indicates the instant $t=4415.7\,\mathrm{s}$ when the principal X-point moves from the chromosphere into the corona, and the second vertical dashed line indicates the instant $t=4544.5\,\mathrm{s}$ when the soft X-ray intensity starts to increase.\label{fig4}}
\end{figure*}

In a three-dimensional system, the reconnection rate is equal to the integral of the electric field along the magnetic null line \citep{PriestForbes2000}. The magnetic null line is parallel to the {\it z}-direction in this 2.5 dimensional system, along which the system is translationally invariant. Thus, we only consider the reconnection in the {\it x-y} plane. 
Now we consider the global reconnection rate, which is defined as the sum of the values of the magnetic fluxes transported across all X-points per unit time \citep{Priest2016ASSL}, i.e.,
\begin{equation}
E_{g}=\sum_{X_{p}}\left |-\frac{1}{c} \frac{\mathrm{d} A(X_{p})}{\mathrm{d} t} \right |,\label{eq4}
\end{equation}
where the summation is taken over all X-points. The global reconnection rate is plotted over time (blue curve) in the right panel of Figure \ref{fig4}. The X-point with the maximal reconnection rate among all X-points is called the principal X-point (PX) \citep{Shen2011}, the height of which is plotted over time in the left panel of Figure \ref{fig4}. The reconnection rate at the PX, denoted as $E_{PX}$, is plotted over time (red curve) in the right panel of Figure \ref{fig4}. 

There are three vertical solid lines and two vertical dashed lines in each panel of Figure \ref{fig4}. The first vertical solid line indicates the instant when the resistive instabilities start, i.e., the start of the dynamic growth stage of the CS evolution. The second vertical solid line represents the start of the hot CS stage of the evolution, and the third vertical solid line marks the beginning of the dynamic hot CS stage of the CS evolution. As shown in the left panel of Figure \ref{fig4}, the PX jumps from the chromosphere into the corona at $t=4415.7\,\mathrm{s}$, marked by the first vertical dashed line, before which the PX stays in the chromosphere. The trajectory of the PX is smooth before the first vertical dashed line, while it oscillates after the first vertical dashed line. The trajectory of the PX becomes smooth again in the hot CS stage of the evolution because there are only a few X-points at this stage. In the dynamic hot CS stage, the height of the PX oscillates again because multiple magnetic islands reappear. In the right panel of Figure \ref{fig4}, the global reconnection rate (blue line) and the reconnection rate of the PX (red line) coincide in the CS growth stage and the hot CS stage of the CS evolution, while the global reconnection rate is larger than the reconnection rate of the PX in the dynamic growth stage and the dynamic hot CS stage of the CS evolution. In the CS growth stage of the CS evolution, the time profile of the reconnection rate is smooth, and it starts to oscillate once the resistive instabilities are invoked, i.e., after the first vertical solid line. Both the global reconnection rate and the reconnection rate at the PX start to increase once the resistive instabilities start, and the reconnection rates drastically increase after the first vertical dashed line that marks the instant when the PX jumps from the chromosphere into the corona. We see from the right panel of Figure \ref{fig4} that the reconnection rate reaches one of its maximum values at the second vertical dashed line, i.e., at $t=4544.5\,\mathrm{s}$. Then, the reconnection rate declines abruptly to a low value at the second vertical solid line. The reconnection rate increases drastically when the dynamic hot CS stage of the evolution begins, accompanied by the reappearance of multiple magnetic islands.

\section{Forward-modeling Technique   }\label{sec:3}

The main task of the forward-modeling analysis is to calculate the radiation intensity by solving the radiative transfer equation. The time-independent version of the radiative transfer equation reads \citep{Rybicki1986}
\begin{equation}
\frac{ \mathrm{d} I_{\lambda} }{\mathrm{d} \tau_{\lambda}}=-I_{\lambda}+S_{\lambda}.\label{eq5}
\end{equation}
Here, $I_{\lambda}$ is the specific radiation intensity at wavelength $\lambda$, $\tau_{\lambda}$ is the optical depth, and $S_{\lambda}$ is the source function. The optical depth $\tau_{\lambda}$ is defined by $\mathrm{d}\tau_{\lambda}=\alpha_{\lambda}\mathrm{d}s$, where $\mathrm{d}s$ is a differential element of length along the ray, and $\alpha_{\lambda}$ is the absorption coefficient at wavelength $\lambda$. The source function $S_{\lambda}$ is defined as the ratio of the emission coefficient $j_{\lambda}$ to the absorption coefficient $\alpha_{\lambda}$, i.e., $S_{\lambda}=j_{\lambda}/\alpha_{\lambda}$. The formal solution of the transfer equation is as follows \citep{Rybicki1986}:

\begin{equation}
I_{\lambda}(\tau_{\lambda})=I_{\lambda}(0)\exp(-\tau_{\lambda}) +\int_{0}^{\tau_{\lambda}}\exp[-(\tau_{\lambda}-\tau_{\lambda}^{\prime})] S_{\lambda}(\tau_{\lambda}^{\prime})\mathrm{d}\tau_{\lambda}^{\prime}.\label{eq6}
\end{equation}

In this work, the line-of-sight integral in the radiative transfer equation is taken along our invariant {\it z}-axis, along which the source function is constant because the system is translationally invariant due to the 2.5 dimensional setup.
The formal solution of the radiative transfer equation with a constant source function is as follows \citep{Rybicki1986}:  
\begin{equation}
I_{\lambda}(\tau_{\lambda})=I_{\lambda}(0)\exp(-\tau_{\lambda}) +S_{\lambda}[1-\exp(-\tau_{\lambda})].\label{eq7}
\end{equation}
For the optically thin case, i.e. in the limiting case of $\tau_{\lambda}\to 0$, we have $\exp(-\tau_{\lambda})= 1-\tau_{\lambda}+O(\tau_{\lambda}^{2})$, then Equation (\ref{eq7}) is reduced to 
\begin{equation}
I_{\lambda}=\lim_{\tau_{\lambda}\to 0} I_{\lambda}(0)\exp(-\tau_{\lambda}) +\lim_{\tau_{\lambda}\to 0}\frac{j_{\lambda}}{\alpha_{\lambda}}[\tau_{\lambda}+O(\tau_{\lambda}^{2})]=I_{\lambda}(0) +j_{\lambda}L, \label{eq8}
\end{equation}
with $L$ denoting the length of the ray.

We assume that the absorption for EUV radiation is due to photoionization while the thermal X-ray radiation is considered to be optically thin. Thus, the radiation intensities of the seven channels of \textit{SDO}/AIA, i.e., $94\,\mathrm{\AA}$, $131\,\mathrm{\AA}$, $171\,\mathrm{\AA}$, $193\,\mathrm{\AA}$, $211\,\mathrm{\AA}$, $304\,\mathrm{\AA}$, and $335\,\mathrm{\AA}$, are calculated based on Equation (\ref{eq7}) while the thermal X-ray is obtained by numerically calculating Equation (\ref{eq8}). To calculate Equation (\ref{eq7}) and Equation (\ref{eq8}), we require the emission coefficient $j_{\lambda}$ and the absorption coefficient $\alpha_{\lambda}$. In Section \ref{sec:3.1}, we calculate the electron and ion number density distributions by solving the Saha equation, which are utilized for calculate the emission and absorption coefficients. The monochromatic emission coefficient for EUV emission is estimated in Section \ref{sec:3.2}. The photoionization absorption coefficient for EUV radiation is given in Section \ref{sec:3.3}. In Section \ref{sec34}, we discuss to what extent the photoionization absorption affects the result and the role of the first term on the right-hand side of Equation (\ref{eq7}), i.e. $I_{\lambda}(0)$. The thermal X-ray emission is estimated in Section \ref{sec:3.5}. It should be noted that the above method goes beyond what \textit{FoMo} does, because \textit{FoMo} just deals with optically thin plasma.

\subsection{Calculating the Number Density by Solving the Saha Equation  }\label{sec:3.1}

If the coronal and chromospheric plasmas are supposed to be in local thermal equilibrium, then the relative populations of the various atomic levels are obtained by solving the Saha equation based on the temperature and density obtained from the MHD simulation. To solve the Saha equation, we assume that the coronal and chromospheric plasmas are composed of hydrogen and helium with a $10:1$ abundance. The atomic partition functions in the Saha equation are calculated following \cite{Cardona2010}, and the statistical weights and the ionization potentials for the ground states of atoms are given in \cite{Dappen2000}. 

\begin{figure*}[ht!]
	\epsscale{1.25}
	\plotone{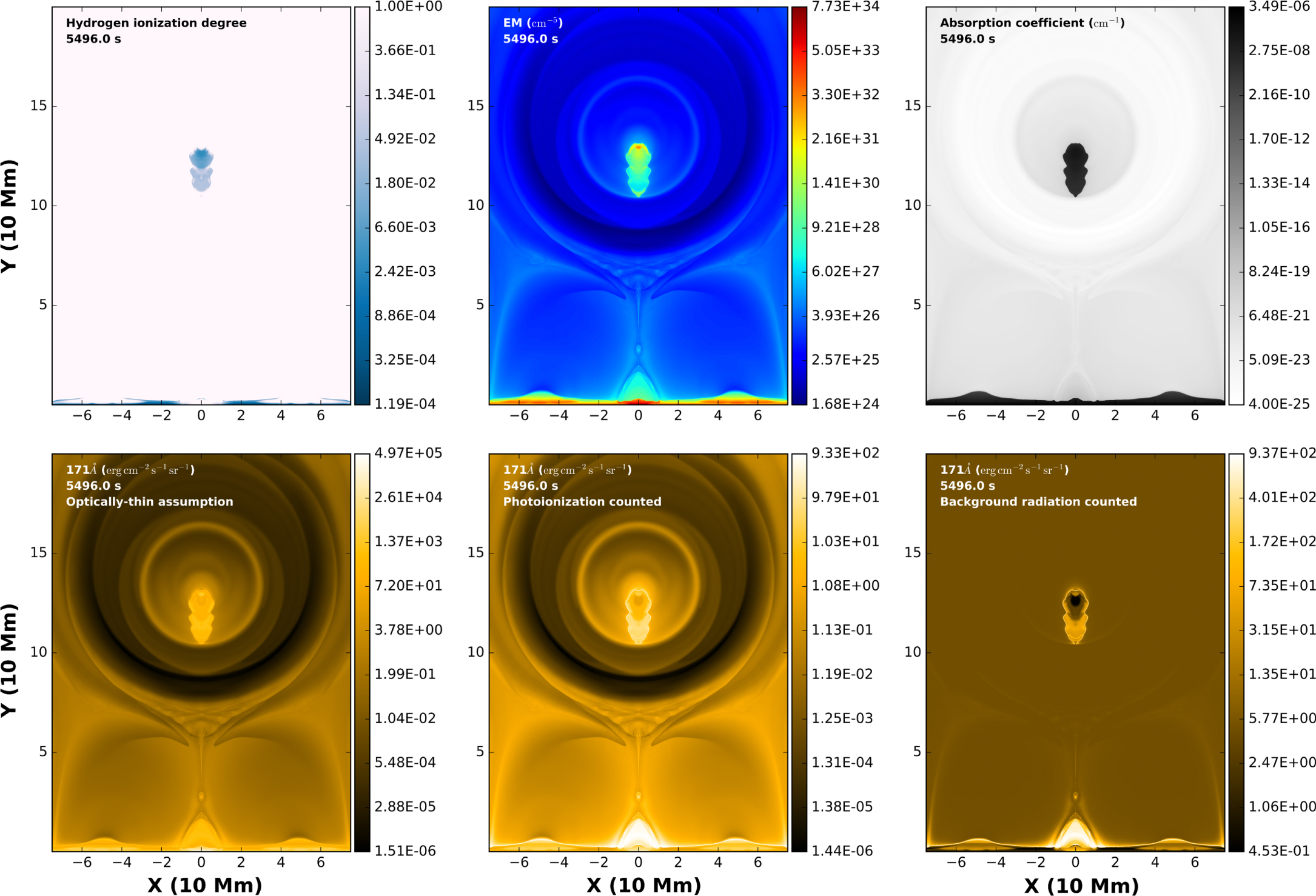}  
	
	\caption{The hydrogen ionization degree is plotted in the upper left panel; the electron number density squared multiplied by the length of the ray, i.e. the so-called emission measure (EM), is plotted in the upper middle panel; the absorption coefficient is plotted in the upper right panel; the radiation intensity of the \textit{SDO}/AIA $171\,\AA$ channel under the optically thin assumption is plotted in the bottom left panel; the radiation intensity of the \textit{SDO}/AIA $171\,\AA$ channel with the photoionization effect accounted for is plotted in the bottom middle panel; and the radiation intensity of the \textit{SDO}/AIA $171\,\AA$ channel with a background radiation, which is $5\%$ of the highest radiation, is plotted in the bottom right panel.\label{fig5}}
\end{figure*}

The hydrogen ionization degree, i.e. the ratio of the ionized hydrogen and the total hydrogen, is plotted in the upper left panel of Figure \ref{fig5}. In the solar corona, the hydrogen is almost fully ionized while in the chromosphere and prominence, only $0.01\%-1\%$ of the hydrogen is ionized, indicating that the prominence and chromosphere are made up of low-temperature weakly ionized plasmas as expected. The upper middle panel of Figure \ref{fig5} shows the emission measure (EM) distribution, which is defined as the free electron number density squared, integrated along the line of sight, i.e. $\mathrm{EM}=\int n_{e}^{2} \mathrm{d}l$, with the length of the ray of the integral $L=100\,\mathrm{Mm}$. The EM distribution indicates that the EM is higher in the prominence and chromosphere even though the ionization degree is small in the chromosphere and prominence, due to the relatively larger number density of atoms in the prominence and chromosphere than in the corona.

\subsection{Estimating the Monochromatic Emission Coefficient for EUV Emission  }\label{sec:3.2}

The monochromatic emission coefficient, proportional to the square of the electron number density $n_{e}$, is calculated as follows \citep{VanDoorsselaere2016}:
\begin{equation}
j_{\lambda}(\mathbf{r})=\frac{A_{b}}{4\pi}n_{e}^{2}(\mathbf{r})G_{\lambda}(n_{e}(\mathbf{r}),T(\mathbf{r})).\label{eq9}
\end{equation}
The abundance of the emitting element $A_{b}$ and the contribution function $G_{\lambda_{0}}$ for the specific spectral line are read from the CHIANTI database \citep{Dere1997,Landi2013}.  

\subsection{Estimating the Continuum Absorption Coefficient for EUV Emission }\label{sec:3.3} 

Cool and dense plasma is considered to be optically thick to EUV radiation, where absorption is due to photoionization of neutral hydrogen and neutral and once-ionized helium \citep{Kucera2015ASSL}. We calculate the absorbing cross section following \citet{Keady2000}, which is also used by \citet{Gibson2016}.

The upper right panel of Figure \ref{fig5} shows the photoionization absorption coefficient distribution at $t=5496.0\,\mathrm{s}$ in the {\it x-y} plane. The photoionization absorption coefficient is $\sim10^{-7}\,\mathrm{cm^{-1}}$ in the prominence and chromosphere while it is less than $\sim10^{-21}\,\mathrm{cm^{-1}}$ in the corona. Suppose the line-of-sight length is $100\,\mathrm{Mm}$, the optical depth is $10^{3}\gg 1$ in the chromosphere and prominence, and $10^{-11}\ll 1$ in the corona, respectively, indicating that the chromosphere and prominence are optically thick while the corona is optically thin.

\subsection{Dark Filament or Bright Prominence} \label{sec34}

\begin{figure*} 
	\epsscale{1.25}
	\plotone{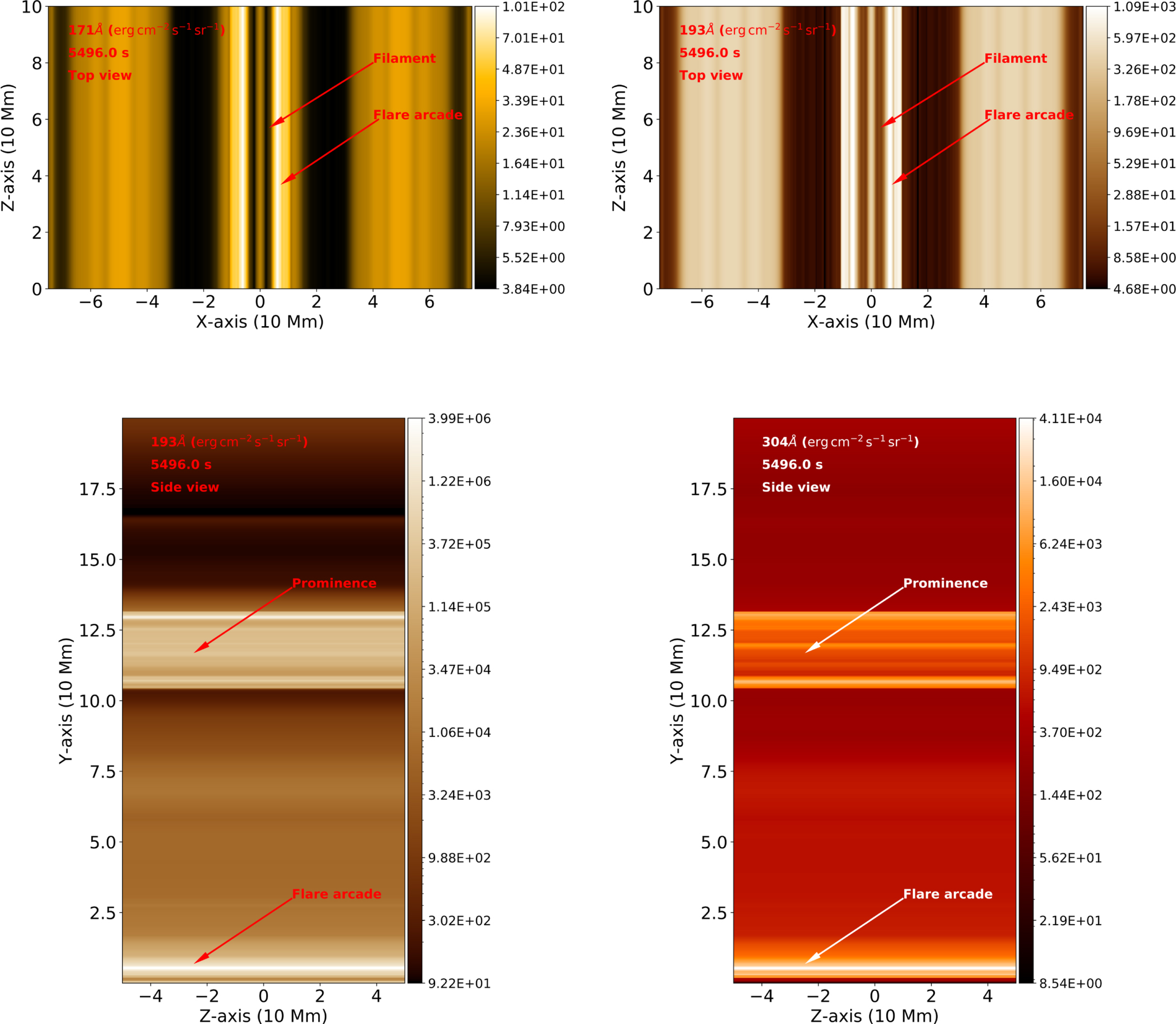}  
	
	\caption{The upper panels show the synthesized \textit{SDO}/AIA $171\,\mathrm{\AA}$ and $193\,\mathrm{\AA}$ images in the {\it x-z} plane viewed from the top, with the dark filament lying on the solar disk. The lower panels show the synthesized \textit{SDO}/AIA $193\,\mathrm{\AA}$ and $304\,\mathrm{\AA}$ images in the {\it z-y} plane viewed from the side.\label{fig6}}
\end{figure*}

The radiation intensities of the seven channels of \textit{SDO}/AIA, i.e., $94\,\mathrm{\AA}$, $131\,\mathrm{\AA}$, $171\,\mathrm{\AA}$, $193\,\mathrm{\AA}$, $211\,\mathrm{\AA}$, $304\,\mathrm{\AA}$, and $335\,\mathrm{\AA}$, can be calculated by using the emission coefficient and absorption coefficient discussed above. In this Section, three cases of forward modeling of EUV emission are investigated in order to clarify the effects of absorption and background radiation. The background radiation refers to $I_{\lambda}(0)$, which appears in the first term on the right-hand side of Equation (\ref{eq7}). The background radiation comes from plasma that is located farther behind the integration depth of $100\,\mathrm{Mm}$. In a realistic solar event, the background radiation may come from the background solar corona.

In Case A, the plasma is assumed to be optically thin, and the background radiation is supposed to be zero, i.e., $\alpha_{\lambda}\to0$ and $I_{\lambda}(0)=0$. Then, the radiation intensity of Case A is calculated by integrating the emission coefficient along the ray, i.e., $I_{\lambda}=j_{\lambda}L$. The radiation intensity for the channel of \textit{SDO}/AIA $171\,\mathrm{\AA}$ in the region $[-75\,\mathrm{Mm},75\,\mathrm{Mm}]\times[0\,\mathrm{Mm} ,200\,\mathrm{Mm}]$ at $t=5496.0\,\mathrm{s}$ is plotted in the bottom left panel of Figure \ref{fig5}. In Case B, the background radiation is still supposed to be zero, and the photoionization effect is taken into account, i.e., $\alpha_{\lambda}\neq 0$ and $I_{\lambda}(0)=0$. We plot the radiation intensity of Case B for \textit{SDO}/AIA $171\,\mathrm{\AA}$ channel in the bottom middle panel of Figure \ref{fig5}. In Case C, the photoionization effect is taken into account, and the background radiation is taken as a nonzero value over the region we considered, i.e., $\alpha_{\lambda}\neq 0$ and $I_{\lambda}(0) \neq 0$. The background radiation accounts for $5\%$ of the highest radiation produced by the plasma, i.e., $I_{\lambda}(0)$ in the first term on the right-hand side of Equation (\ref{eq7}) takes a fraction of $5\%$ of the maximum value of the second term on the right-hand side of Equation (\ref{eq7}). The radiation intensity of this case is plotted in the bottom right panel of Figure \ref{fig5}. The length of the ray of the integral is taken as $L=100\,\mathrm{Mm}$ in Case A, Case B, and Case C.

Comparing Case A (bottom left panel) to Case B (bottom middle panel), we notice that the prominence is brighter than the flare loops under the optically thin assumption due to the relatively higher EM in the prominence than in the flare loops while the flare loops appear brighter than the prominence in Case B because the absorption coefficient in the prominence is more than 14 orders of magnitude higher than that in the flare loops. When the background radiation is counted, i.e. Case C (bottom right panel), the inner and upper parts of the prominence appear dark while the rest is relatively bright. In this case, the prominence may be identified as a filament in observations. The reason why the prominence is dark when the background radiation is counted is that background radiation is an additive to the existing radiation in the optically thin region while some background radiation rays are blocked by the optically thick region. 

The brightness of the prominence in \textit{SDO}/AIA observations depends on the EM, the absorption coefficient, and the background radiation. In a realistic solar atmosphere, the prominence is surely optically thick. When the prominence stands on the limb of the solar disk, the prominence should look bright because the background radiation is relatively weak. If the prominence lies on the solar disk, it will be dark and will be identified as a filament. 

In order to show the dark filament lying on the solar disk, the synthesized \textit{SDO}/AIA $171\,\mathrm{\AA}$ and $193\,\mathrm{\AA}$ images in the {\it x-z} plane are plotted in the upper panels of Figure \ref{fig6} from a top view. The radiation distribution is translationally invariant along the {\it z}-axis due to the 2.5 dimensional setup. It should be noted that when viewing from the top, the source function in the radiative transfer equation is no longer a constant. Equation (\ref{eq6}) is used to obtain the radiation intensity and $I_{\lambda}(0)$ is taken as $0$. Viewing from the top, we verify that the filament looks dark on the solar disk as indicated by the red arrows in the upper panels of Figure \ref{fig6}. The flare arcade is bright, forming two channels around the filament. The dark regions next to the flare arcade correspond to the inner region of the MFR surrounding the prominence with a relatively low temperature. The intermediate layer of the MFR is relatively bright because its temperature is relatively high. The outermost part corresponds to the outer layer of the MFR with a relatively low temperature. We notice that the center of the filament is relatively bright because the top border of the filament is bright with a high EM as can be seen in the middle panels of Figure \ref{fig5}. In the lower panels of Figure \ref{fig6}, we plot the the synthesized \textit{SDO}/AIA $193\,\mathrm{\AA}$ and $304\,\mathrm{\AA}$ images in the {\it z-y} plane viewed from the side. The erupting prominence and the flare arcade are denoted by arrows.

\subsection{Estimating the Thermal X-Ray Emission  }\label{sec:3.5}
Assumed to be optically thin, the thermal X-ray emission is estimated based on the spatial distributions of density and temperature obtained from our MHD simulation. The total bremsstrahlung power radiated by fully ionized plasma per unit volume in the frequency range $(\nu,\nu+d\nu)$ is given by \citet{Rybicki1986} as
\begin{equation}
\frac{\mathrm{d}W}{\mathrm{d}V\mathrm{d}t\mathrm{d}\nu}=C T^{-1/2}(\sum_{i} n_{i}Z^{2}_{i})n_{e}\exp{(-h\nu/k_{B}T)}g_{ff}(h\nu,T),\label{eq10}
\end{equation}
\begin{equation}
C=\frac{2^{5}\pi e^{6} }{3m_{e}c^{3}}(\frac{2\pi}{3k_{B}m_{e}})^{1/2},\label{eq11}
\end{equation}
where $e$ is the electron charge, $m_{e}$ is the electron mass, $c$ is the speed of light in vacuum, $k_{B}$ is the Boltzmann constant, $h$ is the Planck constant, $n_{e}$ is the electron number density, $n_{i}$ is the number density of ions of the kind $i$, $Z_{i}$ is the charge number for ions of the kind $i$, and $g_{ff}(h\nu,T)$ is the Gaunt factor. The number density of ions $n_{i}$, which is either the hydrogen number density $n_{H}$ or the helium number density $n_{He}$, can be calculated based on the assumption of a $10:1$ abundance of hydrogen and helium in this simulation.
The coefficient $C$ can be evaluated in cgs units \citep{Rybicki1986,Aschwanden2005,Pinto2015},
\begin{equation}
C=\frac{2^{5}\pi e^{6} }{3m_{e}c^{3}}(\frac{2\pi}{3k_{B}m_{e}})^{1/2} \approx 6.84\times 10^{-38}\,\mathrm{cm^{3}\,erg\,K^{\frac{1}{2}}}.\label{eq12}
\end{equation}

The Gaunt factor $g_{ff}(h\nu,T)$ is calculated using the method described by \citet{Pinto2015} and \citet{Fang2016} as follows:
\begin{equation}
g_{ff}(h\nu,T)= \left\{
\begin{array}{c c}
1, & h\nu \lesssim k_{B}T, \\ 
(\frac{k_{B}T}{h\nu})^{0.4}, & h\nu > k_{B}T.
\end{array} \right. \label{eq13}
\end{equation}

The thermal bremsstrahlung emission coefficient $j_{\nu}$ is then calculated as follows:
\begin{equation}
j_{\nu}=\frac{1}{4\pi}\frac{\mathrm{d}W}{\mathrm{d}V\mathrm{d}t\mathrm{d}\nu}.\label{eq14}
\end{equation}   

\citet{Pinto2015} and \citet{Fang2016} calculated the thermal soft X-ray photon flux at a distance of $R=1\,\mathrm{AU}$, which is
\begin{equation}
F_{\nu}=\frac{1}{4\pi R^{2}h\nu}\frac{\mathrm{d}W}{\mathrm{d}V\mathrm{d}t\mathrm{d}\nu}=\frac{j_{\nu}}{R^{2}h\nu}.\label{eq15}
\end{equation}
In this paper, we calculate the specific intensity instead, which reads $I_{\nu}(\mathbf{r})=j_{\nu}L$. Here, the thermal bremsstrahlung emission is treated as optically thin, and the length of the ray of the integral is $L=100\,\mathrm{Mm}$.

\section{Results of Forward Modeling  }\label{sec:4}

The emission coefficient and the continuum absorption coefficient for EUV radiation as well as the emission coefficient for thermal X-ray radiation have been calculated in Section \ref{sec:3}. The radiation intensity for the EUV radiation is obtained by inserting the monochromatic emission coefficient for EUV emission calculated in Section \ref{sec:3.2} and the photoionization absorption coefficient calculated in Section \ref{sec:3.3} into Equation (\ref{eq7}) with $I_{\lambda}(0)=0$. The thermal X-ray radiation intensity is obtained by substituting the emission coefficient calculated in Section \ref{sec:3.5} into Equation (\ref{eq8}) with $I_{\lambda}(0)=0$. The synthetic EUV and X-ray images and light curves based on the forward modeling are discussed in Section \ref{sec:4.1} and Section \ref{sec:4.2}.

\subsection{ Synthetic Images  }\label{sec:4.1}

\begin{figure*}[ht!]
	
	\epsscale{1.2}
	\plotone{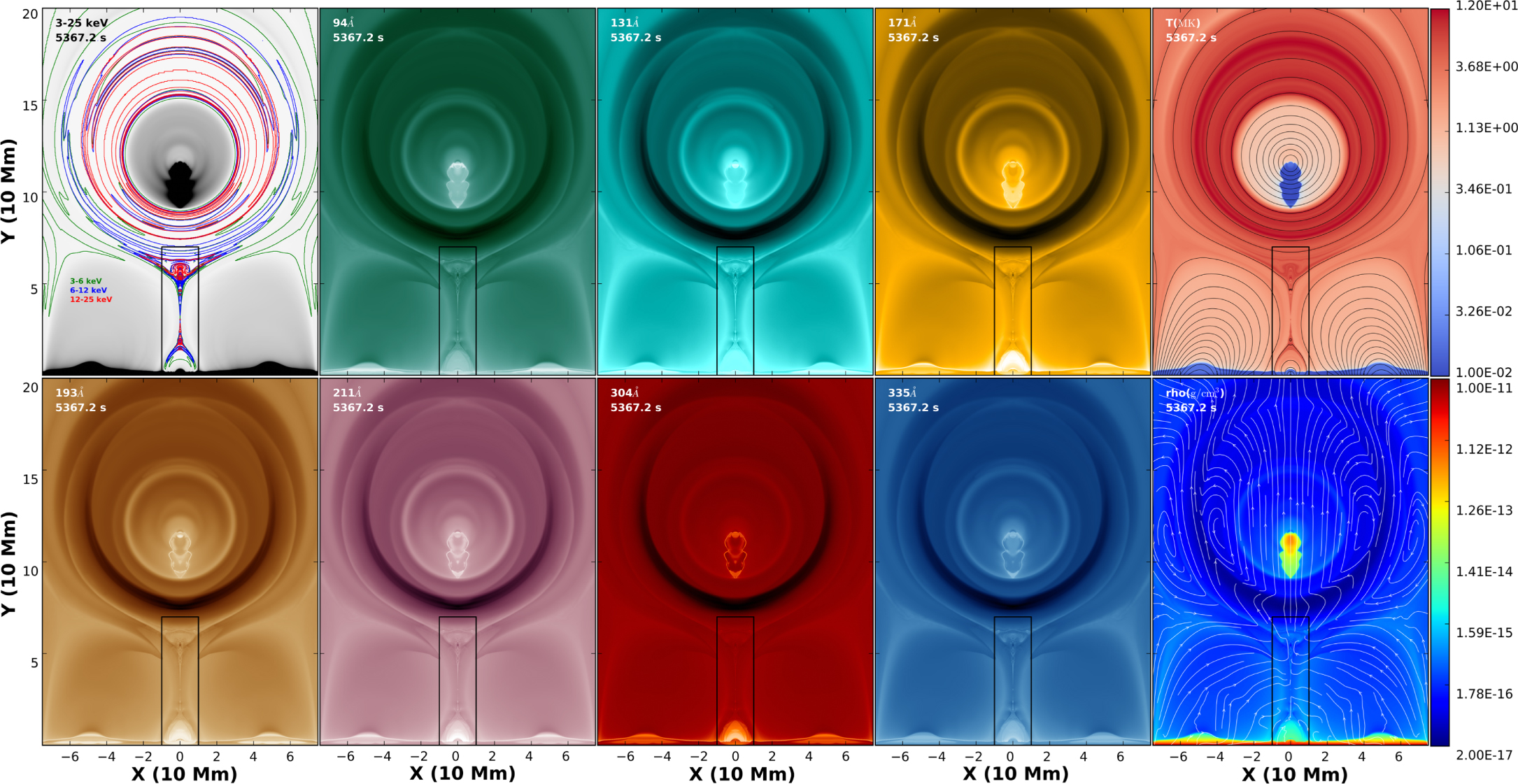} 
\caption{From the upper left to the lower right panel, the $3-25\,\mathrm{keV}$ X-ray intensity with contours of the $3-6\,\mathrm{keV}$ X-ray intensity (green), $6-12\,\mathrm{keV}$ X-ray intensity (blue), and $12-25\,\mathrm{keV}$ X-ray intensity (red); the \textit{SDO}/AIA synthetic views in the $94\,\mathrm{\AA}$, $131\,\mathrm{\AA}$, and $171\,\mathrm{\AA}$ channels; the temperature distribution with magnetic field lines (solid black lines) overlaid; the \textit{SDO}/AIA synthetic views in the $193\,\mathrm{\AA}$, $211\,\mathrm{\AA}$, $304\,\mathrm{\AA}$, and $335\,\mathrm{\AA}$ channels; and the density distribution with velocity vector (white arrows) overlaid at $t=5367.2\,\mathrm{s}$ are plotted. The contours of the $3-6\,\mathrm{keV}$, $6-12\,\mathrm{keV}$, and $12-25\,\mathrm{keV}$ X-ray intensities are identical to those in Figure \ref{fig3}. The color bars of the temperature distribution (upper) and the density distribution (bottom) are plotted on the right of the picture.}\label{fig7} 
\end{figure*}

Synthetic EUV images at $t=5367.2\,\mathrm{s}$ of the seven channels of \textit{SDO}/AIA, i.e., $94\,\mathrm{\AA}$, $131\,\mathrm{\AA}$, $171\,\mathrm{\AA}$, $193\,\mathrm{\AA}$, $211\,\mathrm{\AA}$, $304\,\mathrm{\AA}$, and $335\,\mathrm{\AA}$, are plotted in Figure \ref{fig7} in log scale, together with the temperature distribution (upper right panel) with the magnetic field lines overlaid, the density distribution (bottom right panel) with the velocity streamlines overlaid, and $3-25\,\mathrm{keV}$ X-ray intensity (upper left panel) with contours of $3-6\,\mathrm{keV}$ X-ray intensity, $6-12\,\mathrm{keV}$ X-ray intensity and $12-25\,\mathrm{keV}$ X-ray intensity overlaid. On the right of the picture are plotted the color bars of the temperature distribution (upper) and the density distribution (bottom). The flare arcade is optically thin, the density, close to $\sim2.8\times10^{-15}\,\mathrm{g/cm^{3}}$, and temperature, lying in the range of $2-6\,\mathrm{MK}$, of which are higher than its ambient plasma. In all channels of \textit{SDO}/AIA, there is a dark layer around the MFR due to the relatively low EM there, as discussed in Section \ref{sec34}. The prominence is visible as enhanced emission in the $94\,\mathrm{\AA}$ and $131\,\mathrm{\AA}$ channels, which may be identified as a bright core in these two wavelengths. The temperature that the $171\,\mathrm{\AA}$ channel is sensitive to is around $10^{5.8}\,\mathrm{K}$ \citep{Lemen2012}, close to the flare loop temperature $2-6\,\mathrm{MK}$, thus the flare loops are brighter than the prominence in $171\,\mathrm{\AA}$. Even though the $304\,\mathrm{\AA}$ channel is sensitive to chromosphere and transition region temperatures of about $10^{4.7}\,\mathrm{K}$ \citep{Lemen2012}, the chromosphere and the prominence are darker than the flare loops due to the optically thick effects in these regions. The structures of the MFR and the corona are blurry in $304\,\mathrm{\AA}$ channel. The temperatures that the $193\,\mathrm{\AA}$, $211\,\mathrm{\AA}$, and $335\,\mathrm{\AA}$ channels are sensitive to are about $10^{6.2}\,\mathrm{K}$, $10^{6.3}\,\mathrm{K}$ and $10^{6.4}\,\mathrm{K}$ \citep{Lemen2012}, respectively, suitable for viewing the corona. Thus, the MFR is brighter in these channels than in the $171\,\mathrm{\AA}$ and $304\,\mathrm{\AA}$ channels. From the upper left panel of Figure \ref{fig7}, we see that the thermal X-ray not only comes from the CS but also from the outermost layer and the intermediate layer of the MFR while the inner region of the MFR around the prominence is dark in X-ray channels because the temperature of the outermost and intermediate layers of the MFR is higher than their ambient regions.

\begin{figure*}[ht!]
	
	\epsscale{1.0}
	\plotone{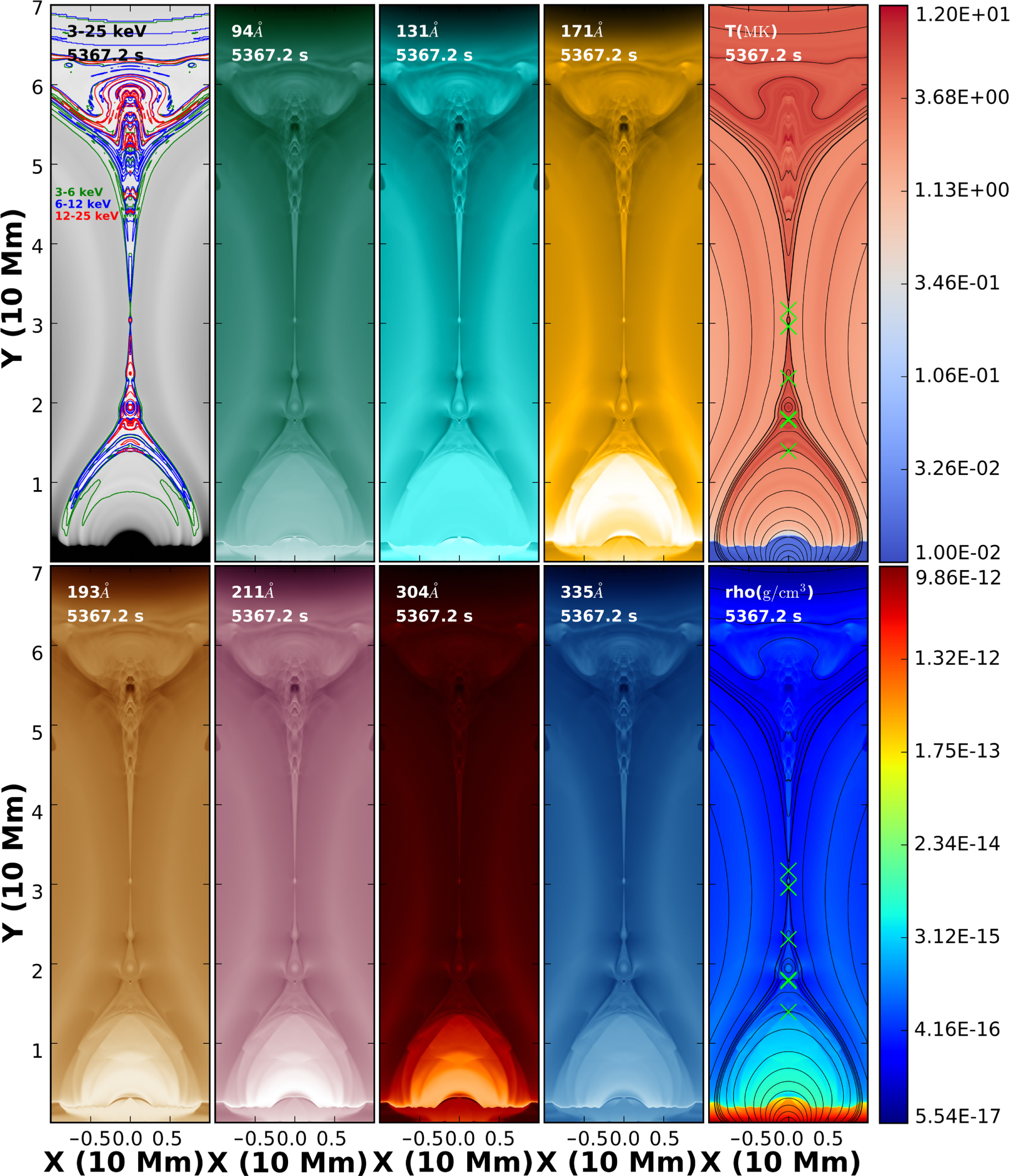}  
	
	\caption{Identical to Figure \ref{fig7} but zooming in on the current sheet structure with magnetic field lines (solid black lines) overlaid on the temperature and density distribution. The green crosses in the temperature and density distribution represent the X-points. The contours of the $3-6\,\mathrm{keV}$, $6-12\,\mathrm{keV}$, and $12-25\,\mathrm{keV}$ X-ray intensities are identical to that in Figure \ref{fig3} and Figure \ref{fig7}. \label{fig8}}
\end{figure*}

Figure \ref{fig8} is a close-up look of the regions inside the black boxes in Figure \ref{fig7}, with a size of $[-10\,\mathrm{Mm},10\,\mathrm{Mm}]\times[0\,\mathrm{Mm} ,70\,\mathrm{Mm}]$, where the CS structure is clearly presented. After the resistive instabilities start, multiple magnetic islands form in the CS. These magnetic  islands of different sizes move bidirectionally upward or downward along the CS, colliding with the flare arcade and the MFR. The motion of the plasma blobs in the CS above the flare arcade has previously been observed by \textit{SDO}/AIA \citep{Takasao2012,Takasao2016}. The region where upward reconnection jets collide with the bottom of the flux rope is referred to as the buffer region in this paper, following \citet{Takahashi2017}. The buffer region is a highly dynamic and complex structure. The Alfv\'enic reconnection jet collides with the buffer region, and more than one shock structure is formed at different heights, which are observable in the seven \textit{SDO}/AIA channels. The buffer region is a layered system consisting of different shock structures. Most shock structures present three components, i.e., two oblique shocks and a horizontal shock in between. \citet{Jelinek2017} studied the oscillations caused by the motion and merging of magnetic islands. \citet{Takasao2015} and \citet{TakasaoShibata2016} showed that waves can be spontaneously generated even if the reconnection outflow is in a quasi-steady state. It is the result of continuous reconnection that higher flare loops have a higher temperature \citep{Hori1997}. Different temperature layers of the flare arcade can be seen from the picture. The inner region of the flare arcade, $2-3\,\mathrm{MK}$, is cooler than the outer layer, $3-7\,\mathrm{MK}$. The outer layer of the arcade is bright in the channel of $131\,\mathrm{\AA}$ while dark in $304\,\mathrm{\AA}$, due to the response ranges of the temperature of the two channels. Owing to the magnetic flux freezing effect and the anisotropy of thermal conduction, the center of the magnetic island is comparatively hot and relatively dense, which leads to a high EM, making it bright in all the channels. There is $12-25\,\mathrm{keV}$ thermal X-ray radiation from the flare loop top and the buffer region of the MFR as shown in the upper left panel in Figure \ref{fig8}. The above-the-loop-top region is full of waves and shocks \citep{TakasaoShibata2016}.

\begin{figure*} 
	
	\epsscale{1.2}
	
	\plotone{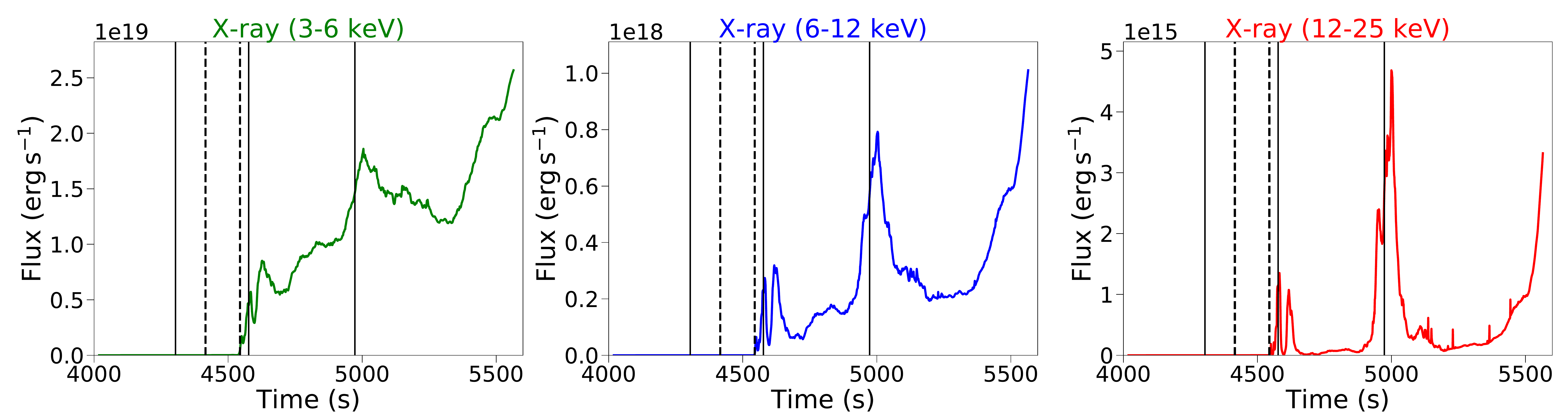} 
	
	\caption{Light curves of the $3-6\,\mathrm{keV}$ X-ray channel (left), the $6-12\,\mathrm{keV}$ X-ray channel (middle), and the $12-25\,\mathrm{keV}$ X-ray channel (right). Vertical lines have the same meaning as in Figure \ref{fig4}.\label{fig9}}
	
\end{figure*}

The evolution of the radiation intensity of the $3-25\,\mathrm{keV}$ X-ray with contours of the $3-6\,\mathrm{keV}$ X-ray intensity (green), the $6-12\,\mathrm{keV}$ X-ray intensity (blue) and the $12-25\,\mathrm{keV}$ X-ray intensity (red) overlaid is plotted in the bottom panels of Figure \ref{fig3}. The CS evolution has four stages in this simulation as discussed in Section \ref{sec:2.2}. The $6-12\,\mathrm{keV}$ and $12-25\,\mathrm{keV}$ radiation does not appear in the CS growth stage of the evolution. The $3-6\,\mathrm{keV}$ radiation is enhanced on the border of the MFR at $t=4293.7\,\mathrm{s}$, a late time of the CS growth stage during the CS evolution, due to the heating of the hot reconnection outflow. The CS and the prominence look dark because their temperatures are relatively low. The resistive instabilities start at $t\sim4304.0\,\mathrm{s}$ and, meanwhile, the dynamic growth stage of the evolution starts. At $4379.6\,\mathrm{s}$, the $3-6\,\mathrm{keV}$ radiation around the border of the MFR increases. At $4465.5\,\mathrm{s}$, the $6-12\,\mathrm{keV}$ and $12-25\,\mathrm{keV}$ radiation appear in the CS and the buffer region of the MFR, where the corresponding temperature rises from $10^{4}\,\mathrm{K}$ to $10^{6}\,\mathrm{K}$ due to the reconnection heating. At $t=4551.4\,\mathrm{s}$, a monster magnetic island appears in the CS after the merging of several magnetic islands. The radiation of the $12-25\,\mathrm{keV}$ channel, which mainly comes from the border around the monster magnetic island, is enhanced at $t=4551.4\,\mathrm{s}$. The hot CS stage of the CS evolution starts at $t\sim4577.1\,\mathrm{s}$, which is indicated by the snapshots at $t=4766.0\,\mathrm{s}$ in Figure \ref{fig3}. In this stage, the prominence has detached from the bottom of the MFR, and the radiation pattern of the thermal X-ray in the buffer region of the MFR is complex and dynamic.

The magnetic islands reappear in the dynamic hot CS stage of the CS evolution, which starts at $t\sim4973.0\,\mathrm{s}$, as indicated by the snapshots at $t=4980.7\,\mathrm{s}$, $t=5195.4\,\mathrm{s}$ and $t=5521.7\,\mathrm{s}$ in Figure \ref{fig3}. At $t=4980.7\,\mathrm{s}$, $12-25\,\mathrm{keV}$ radiation is seen both at the buffer region of the MFR and the top of the flare loops. The $12-25\,\mathrm{keV}$ radiation only appears in the outer layer of the flare loops while the inner part of the flare loops is dominated by the $3-6\,\mathrm{keV}$ and $6-12\,\mathrm{keV}$ radiation because the hot reconnection outflow is confined by the magnetic field to the outer layer of the flare loops, and the inner loops are cooled down by heat conduction and radiative loss. The X-ray radiation continues to increase in the dynamic hot CS stage. At $t=5195.4\,\mathrm{s}$, the flare loops present the cusp-shape structure as discovered by \citet{Masuda1994}. The $12-25\,\mathrm{keV}$ X-ray radiation from the flare loop top and the buffer region of the MFR coincide with the double coronal sources, i.e., the loop-top source and the upper coronal source in soft X-rays, in observations \citep{Sui2003,Ning2016}. We see in this stage that the relatively high-energy radiation, i.e. the $12-25\,\mathrm{keV}$ X-ray radiation, mainly comes from the centers of the magnetic islands while the relatively low-energy radiation originates from the border of the magnetic islands. 
Observational evidence of X-ray plasmoid ejections has been shown by several authors \citep{Shibata1995,Tsuneta1997,Ohyama1998}. \citet{Sun2014} reported X-ray sources that are located between the loop top and the coronal sources within the CS, and they speculated that these sources possibly correspond to the magnetic islands, which is consistent with the results in this work.

It should be noted that chromospheric evaporation is not observed in this simulation. That is because this MHD simulation does not account for the nonthermal electron population that would be accelerated downwards, deposit its energy at chromospheric heights, and cause the evaporation flows. So, the density of the flare arcade is lower than the realistic one, which means that the thermal X-ray and EUV emission from the flare arcade should be weaker than the realistic case. Only thermal X-ray emission is calculated in this work because there nonthermal particles do not exist in this MHD simulation. In the study by \citet{Fang2015,Fang2016} and \citet{Ruan2018arXiv}, an artificial heating function is used to mimic the footpoint heating that triggers chromospheric evaporation. We do not include such a heating function in our simulation either.

\subsection{Light Curves  }\label{sec:4.2}

\begin{figure*}[ht!]
	\epsscale{1.15}
	
	\plotone{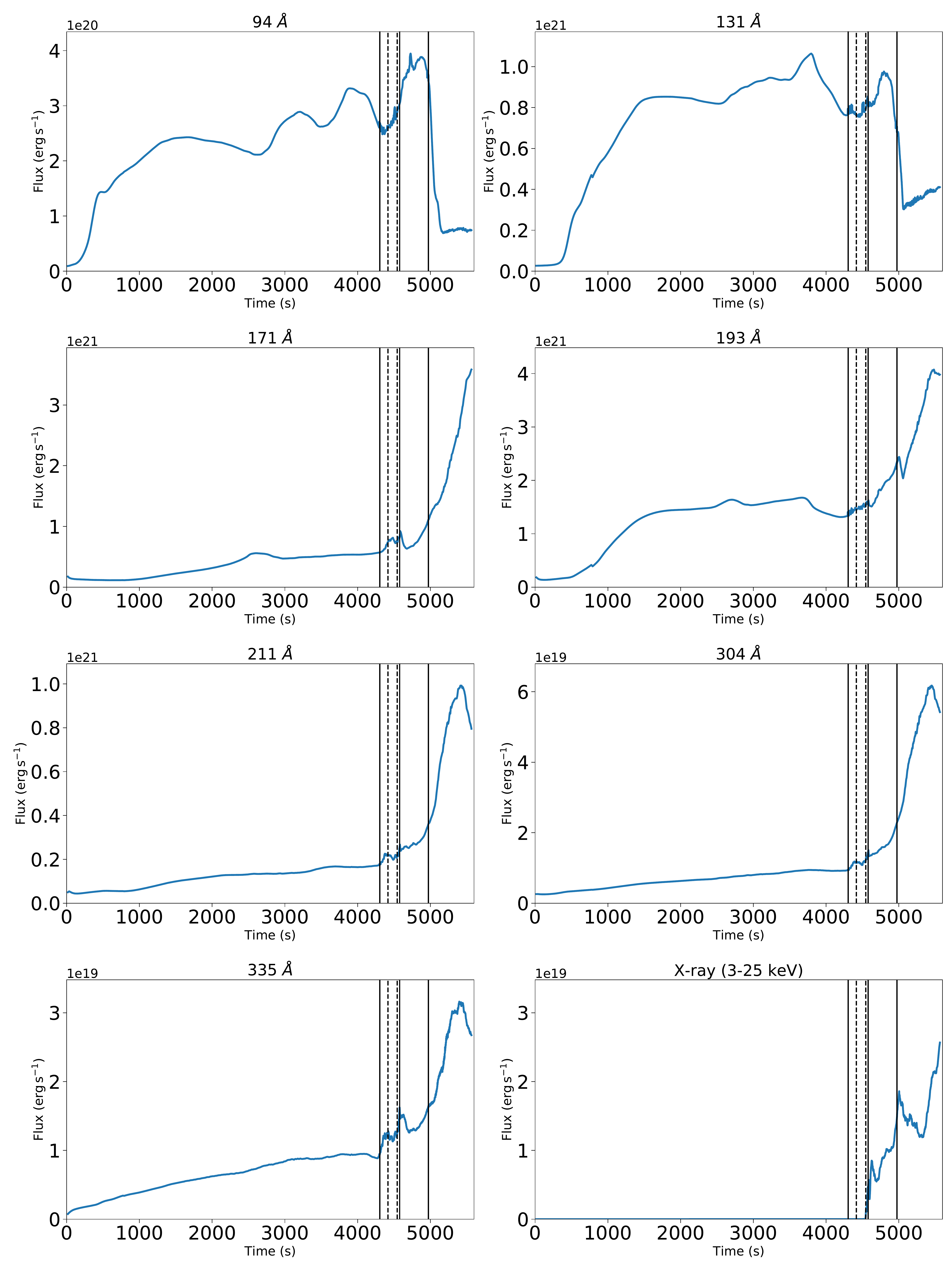} 
	
	\caption{Light curves of the thermal X-ray of $3-25\,\mathrm{keV}$ (bottom right panel) and the seven channels of \textit{SDO}/AIA, i.e., $94\,\mathrm{\AA}$, $131\,\mathrm{\AA}$, $171\,\mathrm{\AA}$, $193\,\mathrm{\AA}$, $211\,\mathrm{\AA}$, $304\,\mathrm{\AA}$, and $335\,\mathrm{\AA}$. Vertical lines have the same meaning as in Figure \ref{fig4} and Figure \ref{fig9}.\label{fig10}}
\end{figure*}

The net flux of energy carried by all photons passing through a certain plane per unit area and per unit time is calculated by integrating the radiation intensity $I$ over the solid angles directed outward of the plane:
\begin{equation}
F=\int I \cos \theta \mathrm{d} \Omega ,\label{eq16}
\end{equation} 
where $\theta$ is the angle subtended between the normal direction of the plane and the differential solid angle $\mathrm{d} \Omega$. Under the assumption of isotropy of radiation intensity, we have
\begin{equation}
F= I\int_{0}^{2\pi}\mathrm{d}\phi\int_{0}^{\pi/2}\cos \theta\sin \theta\mathrm{d}\theta=\pi I.\label{eq17}
\end{equation} 

We focus now on the region $[-10\,\mathrm{Mm},10\,\mathrm{Mm}]\times[0\,\mathrm{Mm},70\,\mathrm{Mm}]$ and calculate the net energy flux $S$ of the EUV radiation of the seven channels of \textit{SDO}/AIA and the thermal X-ray radiation deduced in our simulation as follows,
\begin{equation}
S=\int F \mathrm{d}A= \pi \int I \mathrm{d}A,\label{eq18}
\end{equation} 
where $\mathrm{d}A$ is the differential surface area and the integral is taken over the region $[-10\,\mathrm{Mm},10\,\mathrm{Mm}]\times[0\,\mathrm{Mm},70\,\mathrm{Mm}]$. The radiation intensity is given in Section \ref{sec:3}. The light curves (or time profiles) are obtained by plotting the net flux $S$ over time.

\begin{figure*}[ht!]
	\epsscale{1.2}
	
	\plotone{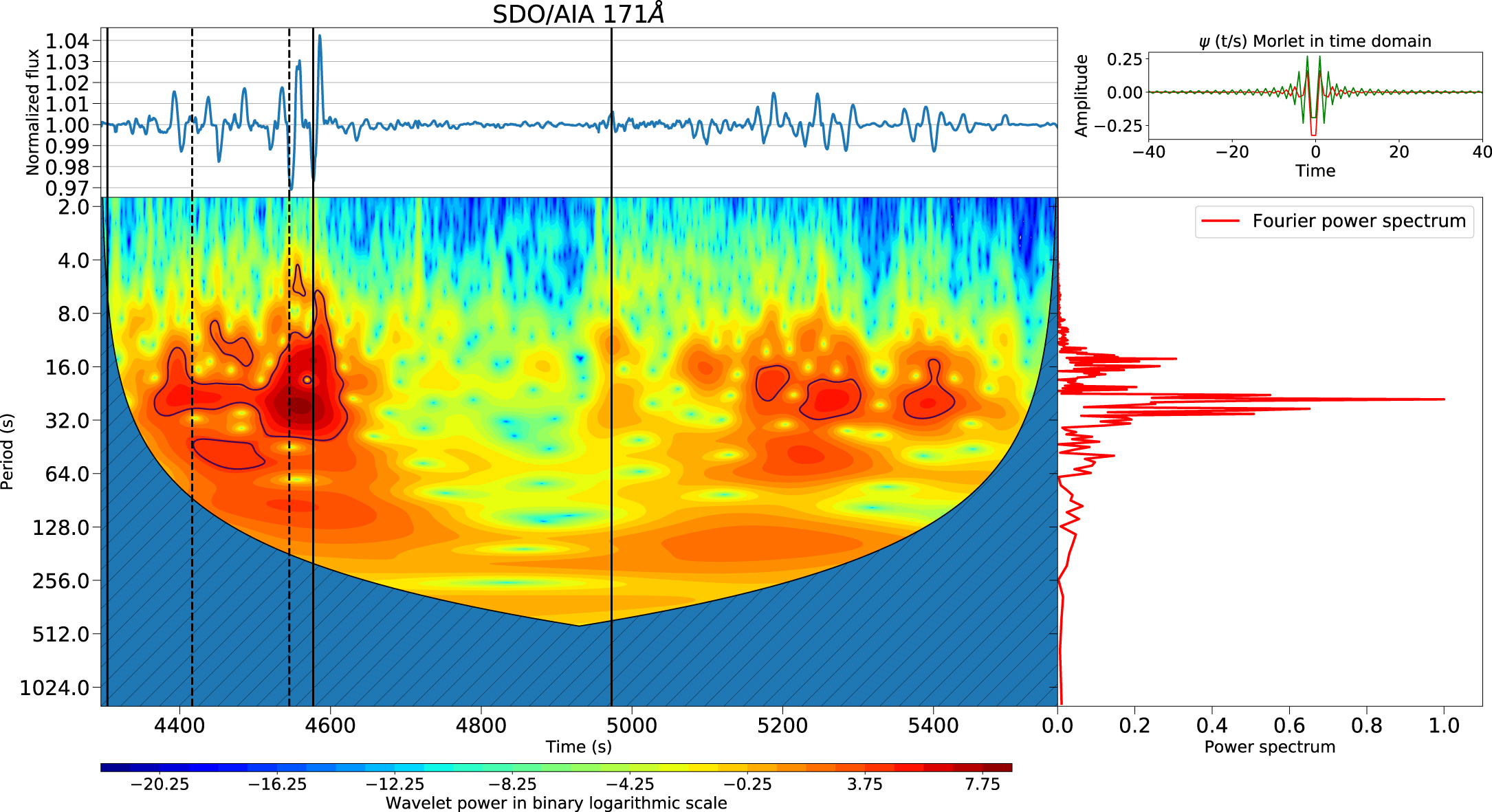} 
	
	\caption{Wavelet transform of the $171\,\mathrm{\AA}$ light curve normalized by robust locally weighted regression. Vertical lines have the same meaning as in Figure \ref{fig4}, Figure \ref{fig9} and Figure \ref{fig10}. \label{fig11}}
\end{figure*}

\begin{table*}[t]
	\begin{center}
		\caption{Quasi-periodic pulsations in different wavelengths and stages.\label{tbl2}}
		\begin{tabular}{r|r|r|r} 
			\tableline\tableline
			\centering
			 .& Dynamic growth stage  & Hot CS stage & Dynamic hot CS stage \\
			\tableline
			\textit{SDO}/AIA $94\,\mathrm{\AA}$  & QPP with a period of $25.7\,\mathrm{s}$ & Weak fluctuation &  Irregular bursty oscillation \\
			\tableline
			\textit{SDO}/AIA $131\,\mathrm{\AA}$  &Irregular oscillation  &Weak fluctuation & Irregular bursty oscillation \\
			\tableline
			\textit{SDO}/AIA $171\,\mathrm{\AA}$  &QPP with a period of $24.1\,\mathrm{s}$  &Weak fluctuation & QPP with a period of $27.3\,\mathrm{s}$  \\
			\tableline
			\textit{SDO}/AIA $193\,\mathrm{\AA}$  &Irregular oscillation  &Weak fluctuation & QPP with a period of $28.5\,\mathrm{s}$ \\
			\tableline
			\textit{SDO}/AIA $211\,\mathrm{\AA}$  &QPP with a period of $29.7\,\mathrm{s}$  &Weak fluctuation & QPP with a period of $14.9\,\mathrm{s}$  \\
			\tableline
			\textit{SDO}/AIA $304\,\mathrm{\AA}$  &QPP with a period of $32.1\,\mathrm{s}$&Weak fluctuation& QPP with a period of $28.5\,\mathrm{s}$  \\
			\tableline
			\textit{SDO}/AIA $335\,\mathrm{\AA}$  &QPP with a period of $25.7\,\mathrm{s}$ &Weak fluctuation &Irregular bursty oscillation  \\
			\tableline
			$3-12\,\mathrm{keV}$ X-ray &Irregular oscillation at the end of the stage  &Weak fluctuation & Weak fluctuation \\
			\tableline
		\end{tabular}
	\end{center}
\end{table*}

Figure \ref{fig9} shows the time profiles of the $3-6\,\mathrm{keV}$ X-ray radiation (left panel), the $6-12\,\mathrm{keV}$ X-ray radiation (middle panel) and the $12-25\,\mathrm{keV}$ X-ray radiation (right panel). The three vertical solid lines indicate the start of the dynamic growth stage of the CS evolution at  $t=4304.0\,\mathrm{s}$, which is also the instant when the resistive instabilities start, the start of the hot CS stage of the evolution at $t=4577.1\,\mathrm{s}$, and the start of the dynamic hot CS stage of the evolution at $t=4973.0\,\mathrm{s}$, respectively. The first vertical dashed line indicates the instant when the PX jumps from the chromosphere into the corona. The second vertical dashed line indicates the instant when the reconnection rate reaches one of its  maximal values. The X-ray radiations of the three channels start to increase after the instant represented by the second vertical dashed line. There are two emission peaks in each channel. The two peaks occur tens of seconds after the second and the third vertical solid lines, respectively, i.e., the two peaks occur successively right after the start of the third and fourth stages of the evolution of the CS. As shown in the right panel of Figure \ref{fig4}, the reconnection rate reaches its maximum value near the second vertical solid line, i.e. $t\sim4577.1\,\mathrm{s}$, then the reconnection rate decreases drastically at $t\sim4577.1\,\mathrm{s}$. During the increase of the global reconnection rate, the energy stored in the magnetic field is released efficiently. A time delay of tens of seconds between the second vertical solid line and the first peak of the X-ray radiation is observed, which is actually the thermal conduction time scale, i.e., the time required for the energy released in the X-points to be conducted outward to heat the CS. After the start of the dynamic hot CS stage of the evolution, the global reconnection rate increases drastically as indicated in the right panel of Figure \ref{fig4}, so the magnetic energy releasing efficiency increases, leading to the second emission peak tens of seconds after the start of the dynamic hot CS stage of the evolution. As shown in Figure \ref{fig3}, the temperature of the ambient coronal plasma around the CS at $t=5521.7\,\mathrm{s}$ is higher than $t=4766.0\,\mathrm{s}$, indicating a higher X-ray emission at a later time in the ambient coronal plasma. Note also that the prominence, which is seen as a dark region with a relatively low X-ray emission, moves out of the box where we calculate the light curves, indicating that the effective X-ray emission area expands with the low-emission prominence replaced by other high-emission structures. These consequences lead to the rise of the light curves in the three X-ray channels after $t\sim5300.0\,\mathrm{s}$, as indicated in Figure \ref{fig9}. The light curves in Figure \ref{fig9} present more bursty structures than observations, which may be attributed to the high spatial and temporal resolution of this simulation. The spatial resolution of the \textit{Reuven Ramaty High-Energy Solar Spectroscopic Imager} (RHESSI) is $\sim2.3\,\mathrm{arcsec}$ \citep{Lin2002SoPh}, which is much more coarse than the resolution of this simulation $24.4\,\mathrm{km}$. The coarse resolution of the instrument is equivalent to the effect of smoothness, leading to the smoothing of some fine structures in images and bursty structures in light curves.

Figure \ref{fig10} indicates the light curves of the thermal X-ray of $3-25\,\mathrm{keV}$ (bottom right panel) and the seven channels of \textit{SDO}/AIA, i.e., $94\,\mathrm{\AA}$, $131\,\mathrm{\AA}$, $171\,\mathrm{\AA}$, $193\,\mathrm{\AA}$, $211\,\mathrm{\AA}$, $304\,\mathrm{\AA}$, and $335\,\mathrm{\AA}$. The vertical lines, both solid and dashed, have the same meaning as those in Figure \ref{fig9}. The light curves of the seven channels of \textit{SDO}/AIA start to rise after the first vertical solid line, i.e., the instant when the resistive instabilities start, then these light curves present some oscillations in the dynamic growth stage of the CS evolution, which is related to the activity of the magnetic islands.

There are two peaks in the thermal X-ray emission, one appears after the second vertical solid line and the other occurs after the third vertical solid line, in the light curve of the thermal X-ray of $3-25\,\mathrm{keV}$ (bottom right panel). Here, we point out that the small emission peaks after the second vertical solid line in the $171\,\mathrm{\AA}$ channel, $193\,\mathrm{\AA}$ channel, and $335\,\mathrm{\AA}$ channel, respectively, have the same origin as the thermal X-ray emission peak after the second vertical solid line while the emission peak after the third vertical solid line in the $193\,\mathrm{\AA}$ channel has the same origin as the emission peak after the third vertical solid line in the thermal X-ray radiation.

The thermal X-ray radiation starts to rise after the instant marked by the second vertical dashed line, i.e., when the global reconnection rate reaches one of its maxima. The radiation of the seven channels of \textit{SDO}/AIA starts to increase when the resistive instabilities are invoked, coinciding with the instant when the transition from the initiation phase to the acceleration phase of MFR occurs, as pointed out by \citet{Zhao2017}. There is a time delay between the increase of the X-ray flux and the increase of the EUV flux, which is close to the time duration of the dynamic growth stage of the CS evolution, about $270\,\mathrm{s}$.

Figure \ref{fig11} shows the normalized light curve of the \textit{SDO}/AIA $171\,\mathrm{\AA}$ channel (upper left panel) after $t=4304.0\,\mathrm{s}$, the wavelet spectrum of the light curve (bottom left panel), the Fourier power spectrum of the normalized light curve (bottom right panel), and the Morlet wavelet that is used for the wavelet transform (upper right panel). The light curve is normalized by a smoothed curve that is obtained by robust locally weighted regression \citep{William1979}, where $1.5\%$ of the total of $1480$ data points are used for computing the value at each point with $50$ iterations applied. The meaning of the vertical lines, both solid and dashed, is the same as that in Figure \ref{fig4}, Figure \ref{fig9}, and Figure \ref{fig10}. The black solid contours in the wavelet spectrum of the light curve (bottom left panel) indicate the $95\%$ confidence level. QPPs appear twice in the normalized light curve (upper left panel). The first QPP behavior appears in the dynamic growth stage of the CS evolution. The amplitude of the first QPP behavior increases in the dynamic growth stage of the CS evolution, especially after the PX jumps from the chromosphere into the corona, and reaches its maximum at the second vertical solid line when the global reconnection rate reaches its maximum. The first QPP behavior disappears after the hot CS stage of the CS evolution begins, during which there is little magnetic island activity, and the global reconnection rate decreases to low values. The wavelet power spectrum (bottom left panel) indicates that the power is enhanced when the global reconnection rate is the highest. The second QPP behavior appears in the dynamic hot CS stage of the CS evolution, which is especially strong between $t=5150\,\mathrm{s}$ and $t=5450\,\mathrm{s}$ and is seen in the wavelet power spectrum (bottom left panel). The Fourier power spectrum (bottom right panel) indicates that the oscillation periods of the two QPP structures are approximately $24.4\,\mathrm{s}$. Applying wavelet analysis to the normalized light curve of the \textit{SDO}/AIA $171\,\mathrm{\AA}$ channel in the dynamic growth stage and the dynamic hot CS stage separately, we notice that the QPP period in the dynamic growth stage is $24.1\,\mathrm{s}$ and the QPP period in the dynamic hot CS stage is $27.3\,\mathrm{s}$, the superposition of the Fourier power spectra in the dynamic growth stage and the dynamic hot CS stage leads to the peak period of $24.4\,\mathrm{s}$ in Figure \ref{fig11}. The origin of the QPP structures comes from the temperature and density oscillations.

It should be noted that the activity of the magnetic islands is intense during the dynamic growth stage and the dynamic hot CS stage of the evolution while the activity of the magnetic islands is weak in the hot CS stage of the evolution as discussed in Section \ref{sec:2.2}. Meanwhile, the appearance of two QPPs coincides with the period of intense activities of the magnetic islands, indicating the close relationship between the two QPPs and the activity of the magnetic islands. The relationship between the activities of the magnetic islands and the QPPs has been investigated in previous studies. \citet{Tan2007} and \citet{Tan2008} suggested that quasiperiodic generations of plasmoids in the CS may lead to QPPs. \citet{Jelinek2017} showed that the motion of magnetic islands may generate waves and shocks that are relevant to QPPs. As listed in Table \ref{tbl2}, the QPPs also appear in other wavelengths. All of the QPPs appear in the dynamic growth stage and the dynamic hot CS stage, and none of the QPPs occurs in the hot CS stage, confirming the relationship between QPPs and the activity of magnetic islands. In the dynamic hot CS stage, QPPs occur in the \textit{SDO}/AIA $171\,\mathrm{\AA}$, $193\,\mathrm{\AA}$, $211\,\mathrm{\AA}$, and $304\,\mathrm{\AA}$ channels because emissions in these channels mainly come from the flare arcade, and QPPs in these wavelengths are related to the flare loop oscillation caused by the hitting of the magnetic islands. The QPP does not appear in the $3-12\,\mathrm{keV}$ soft X-ray channel. The soft X-ray radiation is weak in the dynamic growth stage, and the emission is abruptly enhanced at the end of this stage, leading to irregular oscillation in the normalized light curve at the end of this stage. Thus, the radiation intensity presents weak fluctuation rather than QPP behavior.

\section{Discussion and Conclusion  }\label{sec:5}

In a realistic flare event on the Sun, chromospheric material is heated by precipitating energetic particles accelerated during the flaring process. The heated plasma expands upward and downward. The up-streaming plasma fills up the flare loops, i.e., chromospheric evaporation. In our MHD simulation, nonthermal particles are not taken into account and chromospheric evaporation is not observed. Thus, the density of the postflare arcade is underestimated, which means that the thermal X-ray and the EUV emission from the flare arcade are both underestimated. Another effect due to the absence of nonthermal particles is that there is no hard X-ray emission caused by thick-target bremsstrahlung, which plays an important role in flare observations. However, our model shows that even without thick-target bremsstrahlung, the double X-ray sources, i.e., a loop-top source and a coronal source above the flare arcade, are reproduced. In Section \ref{sec34}, we study three cases in order to clarify the effects of absorption and background radiation in EUV radiation. The brightness of the prominence in \textit{SDO}/AIA observations depends not only on the EM and the absorption coefficient but also on the background radiation.

The main conclusions of the research are summarized as follows.

1. The CS evolution in our MHD simulation of the MFR formation and eruption process driven by photospheric converging motion can be divided into four stages. In the CS growth stage, the CS is formed, the density of which is higher than the ambient coronal plasma while the temperature is lower. There exists only one X-point in the CS in the CS growth stage, which is buried in the chromopshere. The dynamic growth stage starts when resistive instabilities are invoked. The CS is heated from $T\sim10^{4}\,\mathrm{K}$ to $T\sim10^{7}\,\mathrm{K}$ by reconnection, while the density decreases because the magnetic islands carry away mass from the CS. In the hot CS stage, there are few magnetic islands in the CS. In the dynamic hot CS stage, multiple X-points and magnetic islands reappear.

2. We conduct forward-modeling analysis based on our 2.5 dimensional simulation of MFR formation and eruption. The coronal and chromospheric plasma is assumed to be in local thermal equilibrium, then the relative populations of the various atomic levels are obtained by solving the Saha equation based on the temperature and density obtained from the MHD simulation. The EUV emission coefficients of the seven channels of \textit{SDO}/AIA, i.e., $94\,\mathrm{\AA}$, $131\,\mathrm{\AA}$, $171\,\mathrm{\AA}$, $193\,\mathrm{\AA}$, and $211\,\mathrm{\AA}$, are proportional to the electron number density squared. The cool and dense plasma is considered to be optically thick, where the absorption is due to photoionization of neutral hydrogen and neutral and once-ionized helium. Assumed to be optically thin, the thermal X-ray emission is estimated based on the spatial distributions of density and temperature obtained from our MHD simulation. Synthetic images and light curves of the seven channels of \textit{SDO}/AIA and thermal X-ray are obtained.

3. The reconnection rate is relatively low in the CS growth stage of the CS evolution. The reconnection rate starts to increase after the start of the dynamic growth stage of the evolution, accompanied by the rise of the light curves of the seven channels of \textit{SDO}/AIA, i.e., $94\,\mathrm{\AA}$, $131\,\mathrm{\AA}$, $171\,\mathrm{\AA}$, $193\,\mathrm{\AA}$, $211\,\mathrm{\AA}$, $304\,\mathrm{\AA}$, and $335\,\mathrm{\AA}$. The radiation intensity of the $3-25\,\mathrm{keV}$ X-ray starts to increase when the reconnection rate reaches one of its peaks at the end of the dynamic growth stage. The reconnection rate declines to low values in the hot CS stage. The reconnection rate increases drastically again in the dynamic hot CS stage. Accompanied by the increase of the reconnection rate in the dynamic hot CS stage, the second emission peak of the $3-25\,\mathrm{keV}$ X-ray occurs tens of seconds after the start of the dynamic hot CS stage of the evolution. The reconnection rate and the height of the PX oscillate in the dynamic growth stage and the dynamic hot CS stage of the evolution, while they are relatively smooth in the CS growth stage and the hot CS stage.

4. The double soft X-ray sources, i.e., a loop-top source and a coronal source corresponding to the buffer region of the MFR above the flare arcade, are reproduced by forward modeling based on MHD simulation results with the assumption that all X-rays come from thermal emission without a nonthermal component. The magnetic islands in the CS are also X-ray sources.

5. QPPs appear twice in the \textit{SDO}/AIA $171\,\mathrm{\AA}$, $211\,\mathrm{\AA}$, and $304\,\mathrm{\AA}$ channels, one in the dynamic growth stage of the CS evolution, the other in the dynamic hot CS stage, coinciding with the period of intense activities of the magnetic islands, during which the increased magnetic reconnection rate fluctuates significantly over time. QPPs appear once in the \textit{SDO}/AIA $94\,\mathrm{\AA}$ and $335\,\mathrm{\AA}$ channels in the dynamic growth stage, and in the $193\,\mathrm{\AA}$ channel in the dynamic hot CS stage.

\acknowledgments
This work is supported by NSFC grants 11427803 and U1731241, and by CAS Strategic Pioneer Program on Space Science, grant No. XDA15052200. X.Z.Z. acknowledges the Major International Joint Research Project (11820101002) of NSFC. X.Z.Z. also acknowledges the activities of the international science team "Pulsations in solar flares: matching observations and models" supported by the International Space Science Institute-Beijing, China. T.V.D. and R.K. were supported by the GOA-2015-014 (KU Leuven) and the European Research Council (ERC) under the European Union's Horizon 2020 research and innovation programme (grant agreement No. 724326).

\bibliographystyle{apj}

\begin{thebibliography}{}
	\expandafter\ifx\csname natexlab\endcsname\relax\def\natexlab#1{#1}\fi
	
	\bibitem[{{Asai} {et~al.}(2004){Asai}, {Yokoyama}, {Shimojo}, \&
		{Shibata}}]{Asai2004}
	{Asai}, A., {Yokoyama}, T., {Shimojo}, M., \& {Shibata}, K. 2004, \apjl, 605,
	L77
	
	\bibitem[{{Aschwanden}(2005)}]{Aschwanden2005}
	{Aschwanden}, M.~J. 2005, {Physics of the Solar Corona: An Introduction with
		Problems and Solutions}, Astronomy and Planetary Sciences (Springer-Verlag
	Berlin Heidelberg)
	
	\bibitem[{{Biskamp}(1997)}]{Biskamp1997}
	{Biskamp}, D. 1997, {Nonlinear Magnetohydrodynamics} (Cambridge, UK: Cambridge
	University Press)
	
	\bibitem[{{Cardona} {et~al.}(2010){Cardona}, {Mart{\'{\i}}nez-Arroyo}, \&
		{L{\'o}pez-Castillo}}]{Cardona2010}
	{Cardona}, O., {Mart{\'{\i}}nez-Arroyo}, M., \& {L{\'o}pez-Castillo}, M.~A.
	2010, \apj, 711, 239
	
	\bibitem[{{Carmichael}(1964)}]{Carmichael1964}
	{Carmichael}, H. 1964, NASA Special Publication, 50, 451
	
	\bibitem[{{Ciaravella} {et~al.}(2002){Ciaravella}, {Raymond}, {Li}, {Reiser},
		{Gardner}, {Ko}, \& {Fineschi}}]{Ciaravella2002}
	{Ciaravella}, A., {Raymond}, J.~C., {Li}, J., {et~al.} 2002, \apj, 575, 1116
	
	\bibitem[{Cleveland(1979)}]{William1979}
	Cleveland, W.~S. 1979, Journal of the American Statistical Association, 74, 829
	
	\bibitem[{{Cowie} \& {McKee}(1977)}]{Cowie1977ApJ}
	{Cowie}, L.~L., \& {McKee}, C.~F. 1977, \apj, 211, 135
	
	\bibitem[{{D{\"a}ppen}(2000)}]{Dappen2000}
	{D{\"a}ppen}, W. 2000, {Atoms and Molecules}, ed. A.~N. {Cox}, 27. ADS:{http://adsabs.harvard.edu/abs/2000asqu.book...27D}
	
	\bibitem[{{Dere} {et~al.}(1997){Dere}, {Landi}, {Mason}, {Monsignori Fossi}, \&
		{Young}}]{Dere1997}
	{Dere}, K.~P., {Landi}, E., {Mason}, H.~E., {Monsignori Fossi}, B.~C., \&
	{Young}, P.~R. 1997, \aaps, 125, doi:10.1051/aas:1997368
	
	\bibitem[{{Dolla} {et~al.}(2012){Dolla}, {Marqu{\'e}}, {Seaton}, {Van
			Doorsselaere}, {Dominique}, {Berghmans}, {Cabanas}, {De Groof}, {Schmutz},
		{Verdini}, {West}, {Zender}, \& {Zhukov}}]{Dolla2012}
	{Dolla}, L., {Marqu{\'e}}, C., {Seaton}, D.~B., {et~al.} 2012, \apjl, 749, L16
	
	\bibitem[{{Fang} {et~al.}(2015){Fang}, {Yuan}, {Van Doorsselaere}, {Keppens},
		\& {Xia}}]{Fang2015}
	{Fang}, X., {Yuan}, D., {Van Doorsselaere}, T., {Keppens}, R., \& {Xia}, C.
	2015, \apj, 813, 33
	
	\bibitem[{{Fang} {et~al.}(2016){Fang}, {Yuan}, {Xia}, {Van Doorsselaere}, \&
		{Keppens}}]{Fang2016}
	{Fang}, X., {Yuan}, D., {Xia}, C., {Van Doorsselaere}, T., \& {Keppens}, R.
	2016, \apj, 833, 36
	
	\bibitem[{{Fletcher} \& {Hudson}(2008)}]{Fletcher2008}
	{Fletcher}, L., \& {Hudson}, H.~S. 2008, \apj, 675, 1645
	
	\bibitem[{{Forbes} \& {Acton}(1996)}]{ForbesActon1996}
	{Forbes}, T.~G., \& {Acton}, L.~W. 1996, \apj, 459, 330
	
	\bibitem[{{Gibson}(2015)}]{Gibson2015}
	{Gibson}, S. 2015, in IAU Symposium, Vol. 305, Polarimetry, ed. K.~N.
	{Nagendra}, S.~{Bagnulo}, R.~{Centeno}, \& M.~{Jes{\'u}s Mart{\'{\i}}nez
		Gonz{\'a}lez}, 245--250. ADS:{http://adsabs.harvard.edu/abs/2015IAUS..305..245G}
	
	\bibitem[{{Gibson} {et~al.}(2016){Gibson}, {Kucera}, {White}, {Dove}, {Fan},
		{Forland}, {Rachmeler}, {Downs}, \& {Reeves}}]{Gibson2016}
	{Gibson}, S., {Kucera}, T., {White}, S., {et~al.} 2016, Frontiers in Astronomy
	and Space Sciences, 3, 8
	
	\bibitem[{{Gruszecki} {et~al.}(2012){Gruszecki}, {Nakariakov}, \& {Van
			Doorsselaere}}]{Gruszecki2012}
	{Gruszecki}, M., {Nakariakov}, V.~M., \& {Van Doorsselaere}, T. 2012, \aap,
	543, A12
	
	\bibitem[{{Gruszecki} {et~al.}(2011){Gruszecki}, {Vasheghani Farahani},
		{Nakariakov}, \& {Arber}}]{Gruszecki2011}
	{Gruszecki}, M., {Vasheghani Farahani}, S., {Nakariakov}, V.~M., \& {Arber},
	T.~D. 2011, \aap, 531, A63
	
	\bibitem[{Guidoni {et~al.}(2016)Guidoni, DeVore, Karpen, \&
		Lynch}]{Guidoni2016}
	Guidoni, S.~E., DeVore, C.~R., Karpen, J.~T., \& Lynch, B.~J. 2016, The
	Astrophysical Journal, 820, 60
	
	\bibitem[{{Guo} {et~al.}(2013){Guo}, {Bhattacharjee}, \& {Huang}}]{Guo2013}
	{Guo}, L.-J., {Bhattacharjee}, A., \& {Huang}, Y.-M. 2013, \apjl, 771, L14
	
	\bibitem[{{Hirayama}(1974)}]{Hirayama1974}
	{Hirayama}, T. 1974, \solphys, 34, 323
	
	\bibitem[{{Hori} {et~al.}(1997){Hori}, {Yokoyama}, {Kosugi}, \&
		{Shibata}}]{Hori1997}
	{Hori}, K., {Yokoyama}, T., {Kosugi}, T., \& {Shibata}, K. 1997, \apj, 489, 426
	
	\bibitem[{{Hu} \& {Liu}(2000)}]{Hu2000ApJ}
	{Hu}, Y.~Q., \& {Liu}, W. 2000, \apj, 540, 1119
	
	\bibitem[{{Jel{\'{\i}}nek} {et~al.}(2017){Jel{\'{\i}}nek}, {Karlick{\'y}}, {Van
			Doorsselaere}, \& {B{\'a}rta}}]{Jelinek2017}
	{Jel{\'{\i}}nek}, P., {Karlick{\'y}}, M., {Van Doorsselaere}, T., \&
	{B{\'a}rta}, M. 2017, \apj, 847, 98
	
	\bibitem[{{Karlick{\'y}} \& {Kliem}(2010)}]{Karlick2010}
	{Karlick{\'y}}, M., \& {Kliem}, B. 2010, \solphys, 266, 71
	
	\bibitem[{{Keady} \& {Kilcrease}(2000)}]{Keady2000}
	{Keady}, J.~J., \& {Kilcrease}, D.~P. 2000, {Radiation}, ed. A.~N. {Cox (New York: Springer)}. ADS:{http://adsabs.harvard.edu/abs/2000asqu.book...95K}
	
	\bibitem[{{Keppens} {et~al.}(2012){Keppens}, {Meliani}, {van Marle}, {Delmont},
		{Vlasis}, \& {van der Holst}}]{Keppens2012JCoPh}
	{Keppens}, R., {Meliani}, Z., {van Marle}, A.~J., {et~al.} 2012, Journal of
	Computational Physics, 231, 718
	
	\bibitem[{{Ko} {et~al.}(2003){Ko}, {Raymond}, {Lin}, {Lawrence}, {Li}, \&
		{Fludra}}]{Ko2003}
	{Ko}, Y.-K., {Raymond}, J.~C., {Lin}, J., {et~al.} 2003, \apj, 594, 1068
	
	\bibitem[{{Kopp} \& {Pneuman}(1976)}]{KoppPneuman1976}
	{Kopp}, R.~A., \& {Pneuman}, G.~W. 1976, \solphys, 50, 85
	
	\bibitem[{{Kucera}(2015)}]{Kucera2015ASSL}
	{Kucera}, T.~A. 2015, in Astrophysics and Space Science Library, Vol. 415,
	Solar Prominences, ed. J.-C. {Vial} \& O.~{Engvold}, 79. ADS:{http://adsabs.harvard.edu/abs/2015ASSL..415...79K}
	
	\bibitem[{{Landau} \& {Lifshitz}(1975)}]{Landau1975}
	{Landau}, L.~D., \& {Lifshitz}, E.~M. 1975, {The classical theory of fields}. ADS: {http://adsabs.harvard.edu/abs/1975ctf..book.....L}
	
	\bibitem[{{Landi} {et~al.}(2013){Landi}, {Young}, {Dere}, {Del Zanna}, \&
		{Mason}}]{Landi2013}
	{Landi}, E., {Young}, P.~R., {Dere}, K.~P., {Del Zanna}, G., \& {Mason}, H.~E.
	2013, \apj, 763, 86
	
	\bibitem[{{Lemen} {et~al.}(2012){Lemen}, {Title}, {Akin}, {Boerner}, {Chou},
		{Drake}, {Duncan}, {Edwards}, {Friedlaender}, {Heyman}, {Hurlburt}, {Katz},
		{Kushner}, {Levay}, {Lindgren}, {Mathur}, {McFeaters}, {Mitchell}, {Rehse},
		{Schrijver}, {Springer}, {Stern}, {Tarbell}, {Wuelser}, {Wolfson}, {Yanari},
		{Bookbinder}, {Cheimets}, {Caldwell}, {Deluca}, {Gates}, {Golub}, {Park},
		{Podgorski}, {Bush}, {Scherrer}, {Gummin}, {Smith}, {Auker}, {Jerram},
		{Pool}, {Soufli}, {Windt}, {Beardsley}, {Clapp}, {Lang}, \&
		{Waltham}}]{Lemen2012}
	{Lemen}, J.~R., {Title}, A.~M., {Akin}, D.~J., {et~al.} 2012, \solphys, 275, 17
	
	\bibitem[{{Lin} \& {Forbes}(2000)}]{LinForbes2000}
	{Lin}, J., \& {Forbes}, T.~G. 2000, \jgr, 105, 2375
	
	\bibitem[{{Lin} {et~al.}(2002){Lin}, {Dennis}, {Hurford}, {Smith}, {Zehnder},
		{Harvey}, {Curtis}, {Pankow}, {Turin}, {Bester}, {Csillaghy}, {Lewis},
		{Madden}, {van Beek}, {Appleby}, {Raudorf}, {McTiernan}, {Ramaty}, {Schmahl},
		{Schwartz}, {Krucker}, {Abiad}, {Quinn}, {Berg}, {Hashii}, {Sterling},
		{Jackson}, {Pratt}, {Campbell}, {Malone}, {Landis}, {Barrington-Leigh},
		{Slassi-Sennou}, {Cork}, {Clark}, {Amato}, {Orwig}, {Boyle}, {Banks},
		{Shirey}, {Tolbert}, {Zarro}, {Snow}, {Thomsen}, {Henneck}, {McHedlishvili},
		{Ming}, {Fivian}, {Jordan}, {Wanner}, {Crubb}, {Preble}, {Matranga}, {Benz},
		{Hudson}, {Canfield}, {Holman}, {Crannell}, {Kosugi}, {Emslie}, {Vilmer},
		{Brown}, {Johns-Krull}, {Aschwanden}, {Metcalf}, \& {Conway}}]{Lin2002SoPh}
	{Lin}, R.~P., {Dennis}, B.~R., {Hurford}, G.~J., {et~al.} 2002, \solphys, 210,
	3
	
	\bibitem[{{Longcope} \& {Priest}(2007)}]{Longcope2007}
	{Longcope}, D.~W., \& {Priest}, E.~R. 2007, Physics of Plasmas, 14, 122905
	
	\bibitem[{{Masuda} {et~al.}(1994){Masuda}, {Kosugi}, {Hara}, {Tsuneta}, \&
		{Ogawara}}]{Masuda1994}
	{Masuda}, S., {Kosugi}, T., {Hara}, H., {Tsuneta}, S., \& {Ogawara}, Y. 1994,
	\nat, 371, 495
	
	\bibitem[{{McLaughlin} {et~al.}(2018){McLaughlin}, {Nakariakov}, {Dominique},
		{Jel{\'{\i}}nek}, \& {Takasao}}]{McLaughlin2018}
	{McLaughlin}, J.~A., {Nakariakov}, V.~M., {Dominique}, M., {Jel{\'{\i}}nek},
	P., \& {Takasao}, S. 2018, \ssr, 214, 45
	
	\bibitem[{{Mei} {et~al.}(2017){Mei}, {Keppens}, {Roussev}, \& {Lin}}]{Mei2017}
	{Mei}, Z.~X., {Keppens}, R., {Roussev}, I.~I., \& {Lin}, J. 2017, \aap, 604, L7
	
	\bibitem[{{Nakariakov} {et~al.}(2006){Nakariakov}, {Foullon}, {Verwichte}, \&
		{Young}}]{Nakariakov2006}
	{Nakariakov}, V.~M., {Foullon}, C., {Verwichte}, E., \& {Young}, N.~P. 2006,
	\aap, 452, 343
	
	\bibitem[{{Nakariakov} \& {Zimovets}(2011)}]{Nakariakov2011}
	{Nakariakov}, V.~M., \& {Zimovets}, I.~V. 2011, \apjl, 730, L27
	
	\bibitem[{Ni {et~al.}(2012)Ni, Roussev, Lin, \& Ziegler}]{Ni2012}
	Ni, L., Roussev, I.~I., Lin, J., \& Ziegler, U. 2012, The Astrophysical
	Journal, 758, 20
	
	\bibitem[{{Ning} {et~al.}(2016){Ning}, {Li}, \& {Zhang}}]{Ning2016}
	{Ning}, Z., {Li}, D., \& {Zhang}, Q.~M. 2016, \solphys, 291, 1783
	
	\bibitem[{{Nita} {et~al.}(2015){Nita}, {Fleishman}, {Kuznetsov}, {Kontar}, \&
		{Gary}}]{Nita2015}
	{Nita}, G.~M., {Fleishman}, G.~D., {Kuznetsov}, A.~A., {Kontar}, E.~P., \&
	{Gary}, D.~E. 2015, \apj, 799, 236
	
	\bibitem[{{Ohyama} \& {Shibata}(1998)}]{Ohyama1998}
	{Ohyama}, M., \& {Shibata}, K. 1998, \apj, 499, 934
	
	\bibitem[{{Pagano} {et~al.}(2014){Pagano}, {Mackay}, \& {Poedts}}]{Pagano2014}
	{Pagano}, P., {Mackay}, D.~H., \& {Poedts}, S. 2014, \aap, 568, A120
	
	\bibitem[{{Pesnell} {et~al.}(2012){Pesnell}, {Thompson}, \&
		{Chamberlin}}]{Pesnell2012}
	{Pesnell}, W.~D., {Thompson}, B.~J., \& {Chamberlin}, P.~C. 2012, \solphys,
	275, 3
	
	\bibitem[{{Pinto} {et~al.}(2015){Pinto}, {Vilmer}, \& {Brun}}]{Pinto2015}
	{Pinto}, R.~F., {Vilmer}, N., \& {Brun}, A.~S. 2015, \aap, 576, A37
	
	\bibitem[{{Porth} {et~al.}(2014){Porth}, {Xia}, {Hendrix}, {Moschou}, \&
		{Keppens}}]{Porth2014ApJS}
	{Porth}, O., {Xia}, C., {Hendrix}, T., {Moschou}, S.~P., \& {Keppens}, R. 2014,
	\apjs, 214, 4
	
	\bibitem[{{Priest}(2016)}]{Priest2016ASSL}
	{Priest}, E. 2016, in Astrophysics and Space Science Library, Vol. 427,
	Magnetic Reconnection: Concepts and Applications, ed. W.~{Gonzalez} \&
	E.~{Parker}, 101. ADS:{http://adsabs.harvard.edu/abs/2016ASSL..427..101P}
	
	\bibitem[{Priest \& Forbes(2000)}]{PriestForbes2000}
	Priest, E., \& Forbes, T. 2000, Magnetic reconnection. MHD theory and
	applications (Cambridge University Press)
	
	\bibitem[{{Ruan} {et~al.}(2018){Ruan}, {Xia}, \& {Keppens}}]{Ruan2018arXiv}
	{Ruan}, W., {Xia}, C., \& {Keppens}, R. 2018, \aap, 618, A135
	
	\bibitem[{{Russell} \& {Fletcher}(2013)}]{Russell2013}
	{Russell}, A.~J.~B., \& {Fletcher}, L. 2013, \apj, 765, 81
	
	\bibitem[{{Rybicki} \& {Lightman}(1986)}]{Rybicki1986}
	{Rybicki}, G.~B., \& {Lightman}, A.~P. 1986, {Radiative Processes in
		Astrophysics}, 400. ADS:{http://adsabs.harvard.edu/abs/1986rpa..book.....R}
	
	\bibitem[{{Santamaria} {et~al.}(2015){Santamaria}, {Khomenko}, \&
		{Collados}}]{Santamaria2015}
	{Santamaria}, I.~C., {Khomenko}, E., \& {Collados}, M. 2015, \aap, 577, A70
	
	\bibitem[{{Shen} {et~al.}(2011){Shen}, {Lin}, \& {Murphy}}]{Shen2011}
	{Shen}, C., {Lin}, J., \& {Murphy}, N.~A. 2011, \apj, 737, 14
	
	\bibitem[{{Shibata} {et~al.}(1995){Shibata}, {Masuda}, {Shimojo}, {Hara},
		{Yokoyama}, {Tsuneta}, {Kosugi}, \& {Ogawara}}]{Shibata1995}
	{Shibata}, K., {Masuda}, S., {Shimojo}, M., {et~al.} 1995, \apjl, 451, L83
	
	\bibitem[{{Song} {et~al.}(2012){Song}, {Kong}, {Chen}, {Li}, {Li}, {Feng}, \&
		{Xia}}]{Song2012}
	{Song}, H.~Q., {Kong}, X.~L., {Chen}, Y., {et~al.} 2012, \solphys, 276, 261
	
	\bibitem[{{Sturrock}(1966)}]{Sturrock1966}
	{Sturrock}, P.~A. 1966, \nat, 211, 695
	
	\bibitem[{{Sui} \& {Holman}(2003)}]{Sui2003}
	{Sui}, L., \& {Holman}, G.~D. 2003, \apjl, 596, L251
	
	\bibitem[{{Sun} {et~al.}(2014){Sun}, {Cheng}, \& {Ding}}]{Sun2014}
	{Sun}, J.~Q., {Cheng}, X., \& {Ding}, M.~D. 2014, \apj, 786, 73
	
	\bibitem[{{Takahashi} {et~al.}(2017){Takahashi}, {Qiu}, \&
		{Shibata}}]{Takahashi2017}
	{Takahashi}, T., {Qiu}, J., \& {Shibata}, K. 2017, \apj, 848, 102
	
	\bibitem[{{Takasao} {et~al.}(2012){Takasao}, {Asai}, {Isobe}, \&
		{Shibata}}]{Takasao2012}
	{Takasao}, S., {Asai}, A., {Isobe}, H., \& {Shibata}, K. 2012, \apjl, 745, L6
	
	\bibitem[{{Takasao} {et~al.}(2016){Takasao}, {Asai}, {Isobe}, \&
		{Shibata}}]{Takasao2016}
	---. 2016, \apj, 828, 103
	
	\bibitem[{{Takasao} {et~al.}(2015){Takasao}, {Matsumoto}, {Nakamura}, \&
		{Shibata}}]{Takasao2015}
	{Takasao}, S., {Matsumoto}, T., {Nakamura}, N., \& {Shibata}, K. 2015, \apj,
	805, 135
	
	\bibitem[{{Takasao} \& {Shibata}(2016)}]{TakasaoShibata2016}
	{Takasao}, S., \& {Shibata}, K. 2016, \apj, 823, 150
	
	\bibitem[{{Tan}(2008)}]{Tan2008}
	{Tan}, B. 2008, \solphys, 253, 117
	
	\bibitem[{{Tan} {et~al.}(2007){Tan}, {Yan}, {Tan}, \& {Liu}}]{Tan2007}
	{Tan}, B., {Yan}, Y., {Tan}, C., \& {Liu}, Y. 2007, \apj, 671, 964
	
	\bibitem[{{Tsuneta}(1997)}]{Tsuneta1997}
	{Tsuneta}, S. 1997, \apj, 483, 507
	
	\bibitem[{{Van Doorsselaere} {et~al.}(2016{\natexlab{a}}){Van Doorsselaere},
		{Antolin}, {Yuan}, {Reznikova}, \& {Magyar}}]{VanDoorsselaere2016}
	{Van Doorsselaere}, T., {Antolin}, P., {Yuan}, D., {Reznikova}, V., \&
	{Magyar}, N. 2016{\natexlab{a}}, Frontiers in Astronomy and Space Sciences,
	3, 4
	
	\bibitem[{{Van Doorsselaere} {et~al.}(2016{\natexlab{b}}){Van Doorsselaere},
		{Kupriyanova}, \& {Yuan}}]{VanDoorsselaere2016SoPh}
	{Van Doorsselaere}, T., {Kupriyanova}, E.~G., \& {Yuan}, D. 2016{\natexlab{b}},
	\solphys, 291, 3143
	
	\bibitem[{{Xia} \& {Keppens}(2016)}]{Xia2016ApJ}
	{Xia}, C., \& {Keppens}, R. 2016, \apj, 823, 22
	
	\bibitem[{{Xia} {et~al.}(2014){Xia}, {Keppens}, {Antolin}, \&
		{Porth}}]{Xia2014ApJ}
	{Xia}, C., {Keppens}, R., {Antolin}, P., \& {Porth}, O. 2014, \apjl, 792, L38
	
	\bibitem[{{Xia} {et~al.}(2017){Xia}, {Keppens}, \& {Fang}}]{Xia2017AA}
	{Xia}, C., {Keppens}, R., \& {Fang}, X. 2017, \aap, 603, A42
	
	\bibitem[{{Xia} {et~al.}(2018){Xia}, {Teunissen}, {El Mellah}, {Chan{\'e}}, \&
		{Keppens}}]{Xia2018ApJS}
	{Xia}, C., {Teunissen}, J., {El Mellah}, I., {Chan{\'e}}, E., \& {Keppens}, R.
	2018, \apjs, 234, 30
	
	\bibitem[{{Yuan} {et~al.}(2015){Yuan}, {Van Doorsselaere}, {Banerjee}, \&
		{Antolin}}]{Yuan2015}
	{Yuan}, D., {Van Doorsselaere}, T., {Banerjee}, D., \& {Antolin}, P. 2015,
	\apj, 807, 98
	
	\bibitem[{{Zhao} {et~al.}(2017){Zhao}, {Xia}, {Keppens}, \& {Gan}}]{Zhao2017}
	{Zhao}, X., {Xia}, C., {Keppens}, R., \& {Gan}, W. 2017, \apj, 841, 106
	
\end{thebibliography}

\clearpage

\end{CJK}
\end{document}